\documentclass[preprint,sort&compress,12pt]{elsarticle}

\usepackage[top=1in, bottom=1in, left=1in, right=1in]{geometry}

\usepackage{graphicx}
\usepackage{amsmath}
\usepackage{natbib}
\biboptions{sort,compress} 
\usepackage{amssymb}
\usepackage{amsthm}
\usepackage{bbm}
\usepackage{bm}
\usepackage{lineno}
\usepackage{url}
\usepackage{listings}
\usepackage[colorlinks=true]{hyperref}

\usepackage{color,soul}
\usepackage{mathtools}

\usepackage{gensymb}
\usepackage{array}
\usepackage{multirow}
\usepackage{enumerate}
\usepackage{subfig}
\usepackage{algorithm}
\usepackage{algpseudocode}
\usepackage{xcolor}

\usepackage{xfrac}
\usepackage{acronym}
\acrodef{gan}[GAN]{\emph{GAN}}
\acrodef{bc}[BC]{\emph{BC}}
\acrodef{rbc}[RBC]{\emph{RBC}}

\newcommand\Ra{\mbox{\textit{Ra}}}
\newcommand\Prt{\mbox{\textit{Pr}}}
\newcommand\Nu{\mbox{\textit{Nu}}}
\newcommand\Ma{\mbox{\textit{Ma}}}

\newcommand*\patchAmsMathEnvironmentForLineno[1]{%
  \expandafter\let\csname old#1\expandafter\endcsname\csname #1\endcsname
  \expandafter\let\csname oldend#1\expandafter\endcsname\csname end#1\endcsname
  \renewenvironment{#1}%
     {\linenomath\csname old#1\endcsname}%
     {\csname oldend#1\endcsname\endlinenomath}}%
\newcommand*\patchBothAmsMathEnvironmentsForLineno[1]{%
  \patchAmsMathEnvironmentForLineno{#1}%
  \patchAmsMathEnvironmentForLineno{#1*}}%
\AtBeginDocument{%
\patchBothAmsMathEnvironmentsForLineno{equation}%
\patchBothAmsMathEnvironmentsForLineno{align}%
\patchBothAmsMathEnvironmentsForLineno{flalign}%
\patchBothAmsMathEnvironmentsForLineno{alignat}%
\patchBothAmsMathEnvironmentsForLineno{gather}%
\patchBothAmsMathEnvironmentsForLineno{multline}%
}

\newcolumntype{P}[1]{>{\centering\arraybackslash}m{#1}}
\newcolumntype{L}[1]{>{\raggedright\let\newline\\\arraybackslash\hspace{0pt}}m{#1}}
\newcolumntype{C}[1]{>{\centering\let\newline\\\arraybackslash\hspace{0pt}}m{#1}}
\newcolumntype{R}[1]{>{\raggedleft\let\newline\\\arraybackslash\hspace{0pt}}m{#1}}

\begin{document}

\begin{frontmatter}

\title{Enforcing Statistical Constraints in Generative Adversarial Networks for Modeling Chaotic
Dynamical Systems}
\author[vt]{Jin-Long Wu\fnref{fnjl}}
\fntext[fnjl]{Current affiliation: California Institute of Technology, Pasadena, CA 91125, USA.}
\author[lbnl]{Karthik Kashinath}
\author[lbnl]{Adrian Albert}
\author[awi]{Dragos Chirila}
\author[lbnl]{Prabhat}
\author[vt]{Heng Xiao\corref{mycor}}

\cortext[mycor]{Corresponding author}
\ead{hengxiao@vt.edu}

\address[vt]{Kevin T. Crofton Department of Aerospace and Ocean Engineering, Virginia Tech, Blacksburg, VA 24060, USA}
\address[lbnl]{Lawrence Berkeley National Laboratory, Berkeley, CA 94720, USA}
\address[awi]{Alfred Wegener Institute, Helmholtz Centre for Polar and Marine Research (AWI), Bremerhaven, Germany}

\begin{abstract}
Simulating complex physical systems often involves solving partial differential equations (PDEs) with some closures due to the presence of multi-scale physics that cannot be fully resolved. Although the advancement of high performance computing has made resolving small-scale physics possible, such simulations are still very expensive. Therefore, reliable and accurate closure models for the unresolved physics remains an important requirement for many computational physics problems, e.g., turbulence simulation. Recently, several researchers have adopted generative adversarial networks (GANs), a novel paradigm of training machine learning models, to generate solutions of PDEs-governed complex systems without having to numerically solve these PDEs. However, GANs are known to be difficult in training and likely to converge to local minima, where the generated samples do not capture the true statistics of the training data. In this work, we present a statistical constrained generative adversarial network by enforcing constraints of covariance from the training data, which results in an improved machine-learning-based emulator to capture the statistics of the training data generated by solving fully resolved PDEs. We show that such a statistical regularization leads to better performance compared to standard GANs, measured by (1) the constrained model's ability to more faithfully emulate certain physical properties of the system and (2) the significantly reduced (by up to 80\%) training time to reach the solution. We exemplify this approach on the Rayleigh-B\'enard convection, a turbulent flow system that is an idealized model of the Earth's atmosphere. With the growth of high-fidelity simulation databases of physical systems, this work suggests great potential for being an alternative to the explicit modeling of closures or parameterizations for unresolved physics, which are known to be a major source of uncertainty in simulating multi-scale physical systems, e.g., turbulence or Earth's climate.
\end{abstract}

\end{frontmatter}

\setcounter{page}{2}

\section{Introduction}

Complex physical systems are usually characterized by PDE-governed processes with multi-scale physics. Resolving all the scales of multi-scale processes in simulations is still infeasible for many real applications. In practice, simulating those multi-scale physical systems often involves closures to model unresolved processes. However, these closure models also account for major sources of uncertainties in simulation results, partly due to neglecting high-order statistics from the true physical process. With the advancement of high performance computing in the last decades, high-fidelity simulations of complex multi-scale processes such as fully resolved turbulent flow is now becoming available for a growing number of scenarios. More recently, leveraging existing high-fidelity simulation datasets to build models that can emulate complex multi-scale processes (e.g., turbulence or Earth's climate) has been made possible by the rapid growth of the capabilities of machine learning.

Several machine-learning-based approaches have been proposed for improving the simulation of complex PDE-governed systems, e.g., turbulence modeling~\cite{wang17physics-informed,wu2018data,ling2016reynolds}. Machine-learning-based approaches in other complex systems (e.g., weather and climate modeling) is an area of increasing interest~\cite{brenowitz2018prognostic,rasp2018deep,gagne2014machine,dueben2018challenges,schneider2017earth}. In addition to building closure models, machine learning techniques have been applied to various other science and engineering problems that require advanced data analyses of model output, e.g., classification, detection and segmentation of patterns~\cite{liu2016application,racah2017extremeweather,prabhat2017climatenet}. Although machine learning techniques achieved remarkable successes in applications such as image recognition~\cite{lecun2015deep}, the science and engineering communities gradually realize that it is important to inform machine learning some physics instead of merely relying on data-driven discovery for many scientific problems. Ling et al.~\cite{ling2016reynolds} proposed a tensorial neural network to build an objective functional mapping of tensors by embedding a tensor expansion formula into their neural network. Thomas et al.~\cite{thomas2018tensor} built a more general tensor network to ensure the important equivariance by introducing spherical harmonics as filters. Wu et al.~\cite{wu2018data} demonstrated that the important invariance (e.g., Galilean invariance) can be preserved by only using invariants as inputs and outputs for machine learning models. Other researchers incorporated known physics to machine learning techniques as additional constraints. For instance, Karpatne et al.~\cite{karpatne2017theory} incorporated known physical knowledge by adding a penalty term into the machine learning loss function. Rassi et al.~\cite{raissi2019physics,raissi2018hidden} proposed physics-informed neural networks by enforcing the structure of governing equations. Another promising direction is to explore the relations between the machine learning techniques and traditional frameworks for modeling physics, e.g., the analogy between LSTM network and Mori-Zwanzig formalism, for which preliminary success has been demonstrated by Ma et al.~\cite{ma2018model}. In addition, Lusch et al.~\cite{lusch2018deep} demonstrated that the Koopman theory can be made practical by using autoencoder to identify the linear embedding of nonlinear dynamics.

Recently, a novel architecture known as generative adversarial networks (GANs) has been proposed to generate data that mimics certain properties of images or behaviors of a given system. Specifically, GANs consist of a discriminator that learns properties from training samples and a generator that generates samples that mimic the learned properties. Recent research has shown how GANs can be used to generate new solutions of PDE-governed systems by training on existing datasets. For example, King et. al~\cite{king2017creating,king2018deep} used GANs to create turbulent flow realizations and showed that the generated realizations can capture several statistical constraints of turbulent flows, e.g., Komolgorov's $-5/3$ law and small scale intermittency of turbulence. Moreover, GANs have been utilized in extracting information from high-fidelity simulations of other physical systems, e.g., cosmology that involves N-body simulations~\cite{mustafa2017creating}. Although the standard GANs can capture the true distribution of training data when global minimum of the loss function is achieved~\cite{goodfellow2014generative,bousmalis2017unsupervised}, it is well known that GANs can be difficult to train and possibly converges to a local minimum when the complexity of true data distribution increases. Therefore, it is unlikely that standard GANs can capture all the statistics of the solutions for a complex PDE-governed system, indicating that the trained network would be unable to reproduce some important physical and statistical properties of the system. In addition, there are other challenges associated with GANs, e.g., the stability of training~\cite{radford2015unsupervised}, the noise in generated samples~\cite{arjovsky2017wasserstein} and assessment of sample quality~\cite{salimans2016improved}. To improve GANs performances for physical problems, Xie et al.~\cite{xie2018tempogan} incorporated temporal coherence to GANs to generate super-resolution realizations of turbulent flows. Yang et al.~\cite{yang2018physics} encoded into the architecture of GANs the governing physical laws in the form of stochastic differential equations. Stinis et al.~\cite{stinis2018enforcing} augmented the inputs of the discriminator with residuals and introduced noises into training data as inspired by dynamical systems.

Considering the great potential of GANs to physical systems and the limitation of standard GANs, it is worthwhile to investigate a general approach to improve the performance of GANs in emulating physical systems by introducing proper regularization. Inspired by the work of Karpatne et al.~\cite{karpatne2017theory}, we envision to embed both physical constraint (e.g., conservation laws) and statistical constraint (e.g., statistics from the data distribution) into the generator, in order to enhance the robustness and generalization capability of GANs-based physical system emulator. This work focuses on investigating the advantage of incorporating statistical constraints, and the benefits of physical constraints are presented and discussed in a separated work of the authors~\cite{yang2019enforcing}.

In this work, we present a novel approach to improve GANs performance by better preserving high-order statistics when using GANs to build an emulator of complex PDE-governed systems. We tested the statistical constraint with three relatively simple canonical systems, and discussed extensions that are applicable to more complex systems with the ultimate goal of leveraging existing high-fidelity simulation output to emulate unresolved processes, and provide reliable and accurate alternatives to closure models.

\section{Methodology}
\subsection{Generative adversarial networks}

Generative adversarial networks (GANs) have been introduced in \cite{goodfellow2014generative} as a new technique for training machine learning models, originally in the context of computer image synthesis. The original formulation trains models in a purely unsupervised way; yet since their introduction many modifications have been proposed in the literature that make use of available labeled data in a semi-supervised way. The goal of GANs, as initially proposed, is to train a function $G$ that samples from a \textit{complicated, analytically-unknown} distribution of the data that the algorithm sees. 

The main innovation of GANs is the formulation of the training procedure as a zero-sum game between two networks, a generator $G$ and a discriminator $D$. In the standard setting, the generator receives input as an unstructured random noise vector $z$ drawn from a simple distribution (such as an uniform or Gaussian), which it passes through a succession of deconvolutional layers and nonlinear transforms in a deterministic way and outputs a sample $x_\text{fake} = G(z)$. The role of the discriminator $D$ is to act as a classifier, deciding if a sample $x$ it receives is either real or generated by $G$. At optimality (Nash equilibrium in the game between $G$ and $D$), the generator is provably able to produce ``fake'' samples $x_\text{fake}$ that are implicitly drawn from the (unknown) data distribution that $G$ seeks to emulate. 

\begin{figure}[!htbp]
  \centering
  \includegraphics[width=0.9\textwidth]{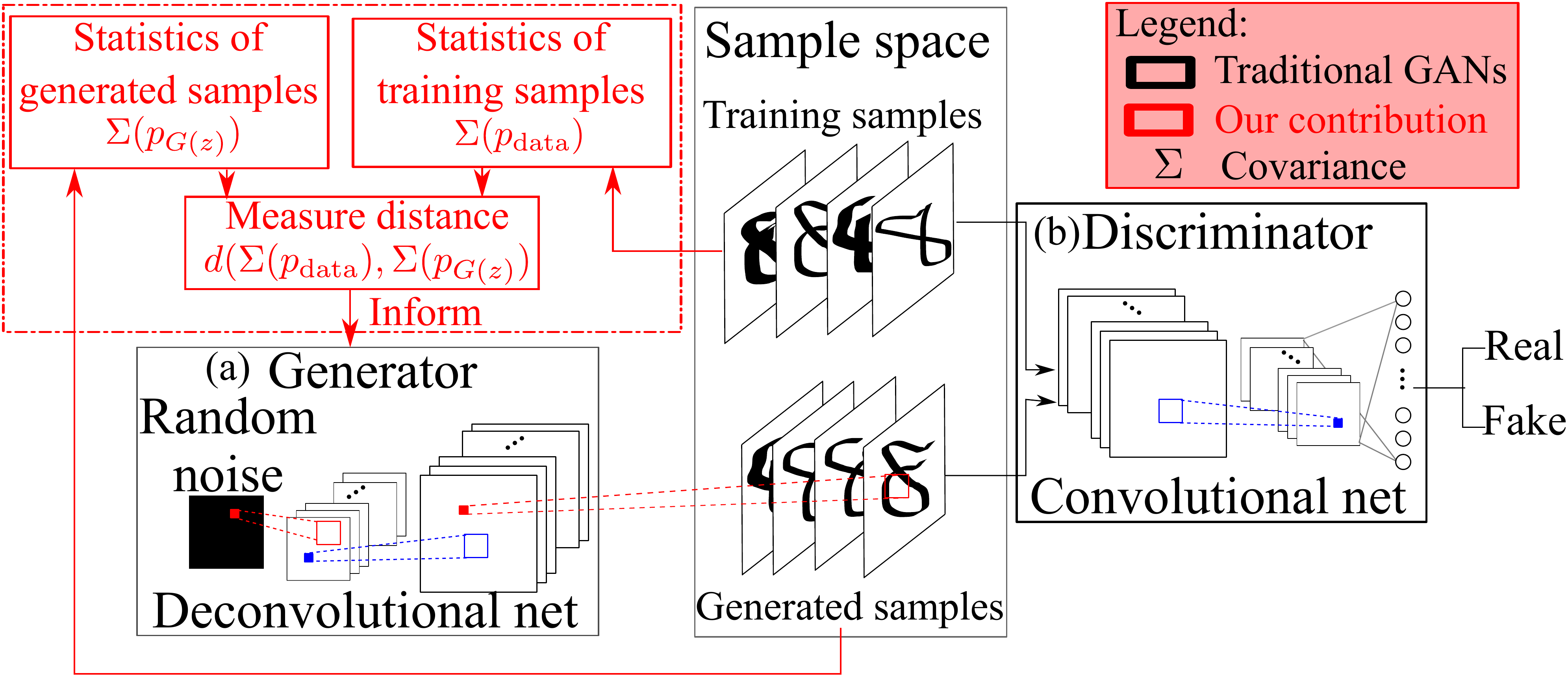}
    \caption{The architecture of a statistics-informed generative adversarial network (GAN), including the architecture of a standard GAN (indicated by the black color) and the modification to help preserving statistics (indicated by the red color).}
  \label{fig:GANs-architecture}
\end{figure}

In this paper, $G$ in Fig.~\ref{fig:GANs-architecture}a and $D$ in Fig.~\ref{fig:GANs-architecture}c are deep deconvolutional neural network and convolutional neural network described by the weights vectors $\theta_G$ and $\theta_D$, respectively. The architectural details follow the structure proposed in \cite{dcgan_2015}. We train the model in the standard way of alternating between the two optimization problems:
\begin{align}
\vspace*{-1em}
\hspace*{-3em}
\theta_D:&\mathcal{L}_D = \mathbb{E}_{x\sim p_{x}} [\text{log }D(x)] + \mathbb{E}_{z\sim p_z}
[\text{log}(1 - D(G(z)))], \\
\theta_G:&\mathcal{L}_G = \mathbb{E}_{z\sim p_z}
[\text{log }(1 - D(G(z)))]
\end{align}
where $\mathbb{E}$ denotes the expectation and $p_{x}$ represents the distribution of the training samples. The training procedure corresponds to a minimax two-player game:
\begin{equation}
\min_{G}\max_{D}L(D,G)
\end{equation}
where the ideal objective is to achieve Nash equilibrium, i.e., the discriminator cannot distinguish the generated samples from the training samples. However, the global optimal is usually difficult to achieve by using standard GANs. Therefore, it is unlikely that standard GANs can identify and preserve all the statistics of the training samples embedded in $p_\textrm{data}$. Those missing statistics are usually important to PDE-governed systems, e.g., the second order moments of the instantaneous velocity corresponds to the Reynolds stress in turbulence modeling. Therefore, we present an approach to better preserve these statistics when using GANs to emulate physical systems.

\subsection{Constrained generative adversarial networks}
The standard formulation above is purely data-driven, i.e., it utilizes no outside knowledge of the problem at hand but that which is encoded implicitly in the data that is used to train the model. As we argued above, constraining the model solution space to physically-feasible regions may yield certain benefits in terms of increased solution quality, decreased training time, and improved data efficiency. Constraints can be imposed in a variety of ways, among which:
\begin{enumerate}[(i)]
\item \emph{Hard constraints on network architecture.} Convolutional networks are by construction translational-invariant and ensure spatial locality of information propagation. It has long been recognized that convolutional filters can be interpreted as discretized differential operators in the finite difference form~\cite{lu2018beyond,latnet_2017,long2018pde}. This fact has recently been exploited to train neural networks as surrogate models of differential equations~\cite{latnet_2017} and to discover PDEs from data~\cite{long2018pde}. Such techniques may be used to constrain the space of feasible solutions of GANs by construction. 
\item \textit{Explicit constraints on optimization.} 
Domain knowledge can be incorporated in a straightforward way through additional penalty terms added into the optimization loss function of GANs. This is in effect a form of regularization that can model many relevant constraints, including conservation laws and high-order statistics. 
\end{enumerate}

In this work, we focus on the second type of constraints discussed above and propose a constrained loss function $L_c(D,G)$ as follow:
\begin{equation}
\label{eq:loss_func}
L_c(D,G)=L(D,G)+\lambda d(\boldsymbol{\Sigma}(p_\textrm{data}),\boldsymbol{\Sigma}(p_{G(z)}))
\end{equation}
where $\boldsymbol{\Sigma}(p)$ denotes the covariance structure of a given distribution $p$, $\lambda$ denotes the coefficient of the penalty term, and $d(\cdot,\cdot)$ represents a distance measure between two covariance structures.  The distance $d$ can be measured in Euclidean space by using Frobenius norm: 
\begin{equation}
F(\boldsymbol{\Sigma}_1,\boldsymbol{\Sigma}_2) = \|\boldsymbol{\Sigma}(p_\textrm{data})-\boldsymbol{\Sigma}(p_{G(z)})\|_F 
\label{eq:frob}
\end{equation}
or by using the symmetrized Kullback--Leibler divergence~\cite{weickert2005visualization}, i.e., 
\begin{equation}
KL_s(\boldsymbol{\Sigma}_1,\boldsymbol{\Sigma}_2)=\frac{1}{2}\textrm{tr}\left(\boldsymbol{\Sigma}_1^{-1}\boldsymbol{\Sigma}_2+\boldsymbol{\Sigma}_2^{-1}\boldsymbol{\Sigma}_1-2\mathbf{I}\right),
\end{equation}
where $\textrm{tr}(\cdot)$ denotes the trace of a matrix, and $\mathbf{I}$ represents the identity matrix. It should be noted that the covariance structure must be positive semidefinite and thus is defined on a low-dimensional manifold within the Euclidean space. Therefore, an alternative is to use Riemannian distance \[
J(\boldsymbol{\Sigma}_1,\boldsymbol{\Sigma}_2)=\sqrt{\sum_{i=1}^{n}\textrm{ln}^2\delta_i(\boldsymbol{\Sigma}_1^{-1}\boldsymbol{\Sigma}_2)}
\] 
where $\delta_i$ denotes the $i$-th eigenvalue of a matrix, or by using the Jensen-Bregman LogDet (JBLD) divergence~\cite{cherian2011efficient}:
\begin{equation}
\label{eq:JBLD}
J(\boldsymbol{\Sigma}_1,\boldsymbol{\Sigma}_2)=\log\left\| \frac{\boldsymbol{\Sigma}_1+\boldsymbol{\Sigma}_2}{2}\right\|-\frac{1}{2}\|\boldsymbol{\Sigma}_1\boldsymbol{\Sigma}_2\|
\end{equation}
We have compared these different choices of the distance measure $d(\boldsymbol{\Sigma}(p_\textrm{data}),\boldsymbol{\Sigma}(p_{G(z)}))$ and concluded that the Frobenius norm serves as the best compromise in terms of the computational cost and the stability of training constrained GANs. The proposed physics-informed GAN is implemented on the machine learning platform \verb+TensorFlow+~\cite{abadi2016tensorflow}. We used a deep convolutional GAN (DCGAN)~\cite{radford2015unsupervised} for most of the  results presented below, but this is still referred to as GAN hereafter for simplicity.

In this work, both the covariance structures of generated and training data are estimated from samples. Specifically, $\Sigma(p_{G(z)})$ is estimated from generated samples at every iteration of training the network. Instead of implicitly estimating the probability distribution of training data $\Sigma(p_\textrm{data})$ as standard GANs, the penalty term in Eq.~\ref{eq:loss_func} explicitly estimates the second-order moment of the distributions and thus better constrains the difference between the distributions of the training data and the generated samples. It should be noted that $\Sigma(p_\textrm{data})$ can also be estimated from the governing equation of the system, since the spatial correlation within the solutions of a PDE-governed system is related to the differential operators of the governing PDEs. More details about this alternative way of estimating the covariance structure of the training data can be found in a separate work~\cite{carlos2019constructing}.

Incorporating physical constraints (either explicitly or implicitly) into GANs may offer certain benefits, of which this paper highlights but several. We posit that appropriately imposing constraints may improve training stability and convergence properties via the regularization these constraints provide. Many hypotheses have been put forth about the source of the training instabilities observed in GANs (mode collapse in particular, as observed initially in \cite{goodfellow2014generative, dcgan_2015}), including different ways in which to measure the distance between the generated and the real data distributions \cite{gulrajani2017improved}, penalties on the gradients around real data points \cite{gan_convergence_stability_2017}, regularizers on the spectral radius of the Jacobian of the gradient vector \cite{numerics_of_gans_2017}, or dynamically choosing the metric to minimize in the objective function via an evolutionary strategy as in \cite{evolutionary_gan_2018}.

\subsection{Lattice Boltzmann Simulation of Rayleigh-B\'enard convection}
\label{sec:LBM}

The physical system we investigated in this work is Rayleigh-B\'enard convection (RBC), which are ultimately driven by buoyancy differences encountered at all
spatial scales, ranging from small engineering devices to Earth sciences and
even astrophysical phenomena. The RBC setup is a classical prototype for
this class of flows. It consists of a slab of fluid, which is bounded by two
horizontal surfaces.\footnote{%
  For simplicity, in analytic or numerical studies the two bounding surfaces are
  sometimes assumed to extend infinitely in the horizontal directions.%
} It was observed from early on \cite{Rayleigh1916} that, when the lower wall is
heated (while the upper wall is cooled), the dynamics of the system is
especially interesting --- a myriad of flow regimes appear, such as stationary
or oscillating convection rolls, convection cells of different shapes, as well
as turbulent flow. Although the detailed dynamics is strongly-dependent on the
initial conditions, the selection of the different regimes is mostly dictated by
the control-parameters, which for the RBC problem are the Rayleigh number
($\Ra = \sfrac{\beta g H^3 (\Delta T)}{\kappa \nu}$ --- the relative magnitude
of the buoyancy velocity to the thermal velocity), the Prandtl number ($\Prt =
\sfrac{\nu}{\kappa}$ --- the ratio of viscous diffusivity to thermal
diffusivity), and the horizontal-to-vertical aspect ratio of the domain ($\Gamma
= \sfrac{W}{H}$). In these definitions, $\beta$ is the coefficient of thermal
expansion, $g$ is the gravitational acceleration, the spatial scales $H$ and $W$
are the height and width respectively, and $\Delta T \ge 0$ is the temperature
difference between the lower and the upper walls.

The dataset we used for training a GAN to emulate the RBC problem
consisted of simulation output, produced with a two-dimensional fluid dynamics
model based on the LBM. The numerical scheme is described in
\cite{Wang2013a} (see \cite{Chirila2018} for a more detailed description). The
values of the control-parameters were $\Prt = 0.71$ (%
for air%
) and $\Ra = 2.5\times 10^8$.
For simplicity (and to facilitate comparison with some theoretical results),
periodic boundary condition (BC) were used at the lateral (vertical) walls, with a relatively
high aspect-ratio of $\Gamma = 7$, which minimizes unphysical artifacts due to
periodicity itself.
The spatial resolution was $\{Nx, Ny\} = \{1792, 256\}$, where the vertical
resolution was the same as used e.g. in \cite{VanderPoel2013} (based on the
theoretical criteria from \cite{Shishkina2010}, to provide for sufficient
resolution within the near-wall boundary layers). The value for the pseudo Mach number (a model-specific parameter) was set to
$\Ma = 0.1$, to keep the artificial compressibility errors below $1\%$.

\begin{figure}[!ht]
  \centering
  \includegraphics[width=1.0\linewidth]{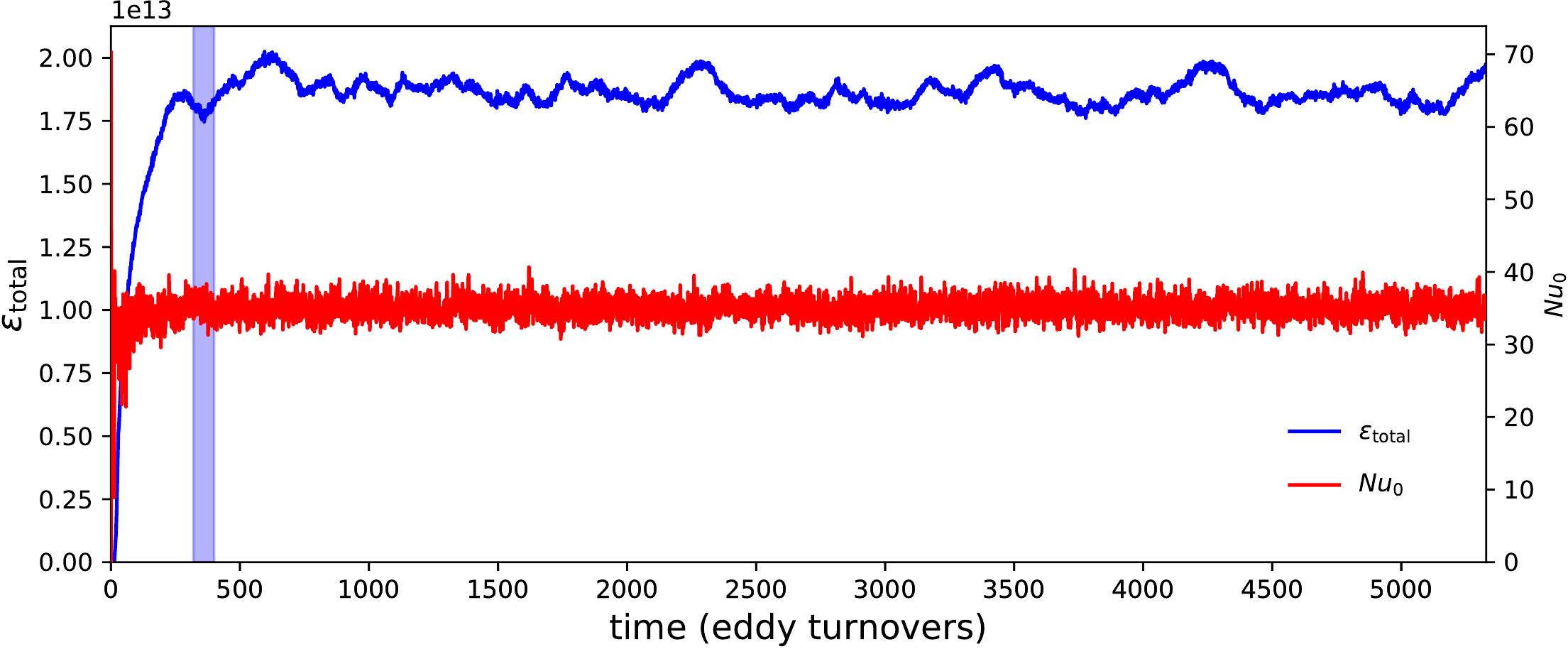}
  \caption{%
    Time-dependence of the total kinetic energy $\varepsilon_\mathrm{total}$
    (blue) and of the Nusselt number $\Nu_{0}$ close to the lower wall (red) for
    the high-resolution \ac{rbc} simulation.
    }
  \label{fig:RB_TKE_and_Nu}
\end{figure}

The model was initially run for 400 eddy turnover times, of which the initial
320 eddy turnover times (necessary for spin-up from the rest state) were
discarded. To verify that the data used for GAN training did in fact
correspond to the final (turbulent) regime, this simulation was later extended
until 5330 eddy turnovers. Two physical metrics of the flow (total kinetic
energy $\varepsilon_\mathrm{total}$ and Nusselt number $\Nu_{0}$ close to the
lower wall) are shown in Figure \ref{fig:RB_TKE_and_Nu}. Both metrics are already
stabilized at the start of the training window, showing that the data used for
GAN training was already part of the final turbulent regime. To further
support this hypothesis of statistical stationarity of the flow, the average
value of the Nusselt number within the training window was found to be
$\langle \Nu_{0} \rangle = 35.1137$, which is close to the values found by other
authors -- for example, \cite{Johnston2009} obtained
$\langle \Nu_{0} \rangle = 34.1422$, for a similar setup, with a lower
aspect-ratio of $\Gamma = 2$ (which artificially constrains the flow,
considering that they also used periodic BC).

\begin{figure}[!ht]
  \centering
  \includegraphics[width=0.8\linewidth]{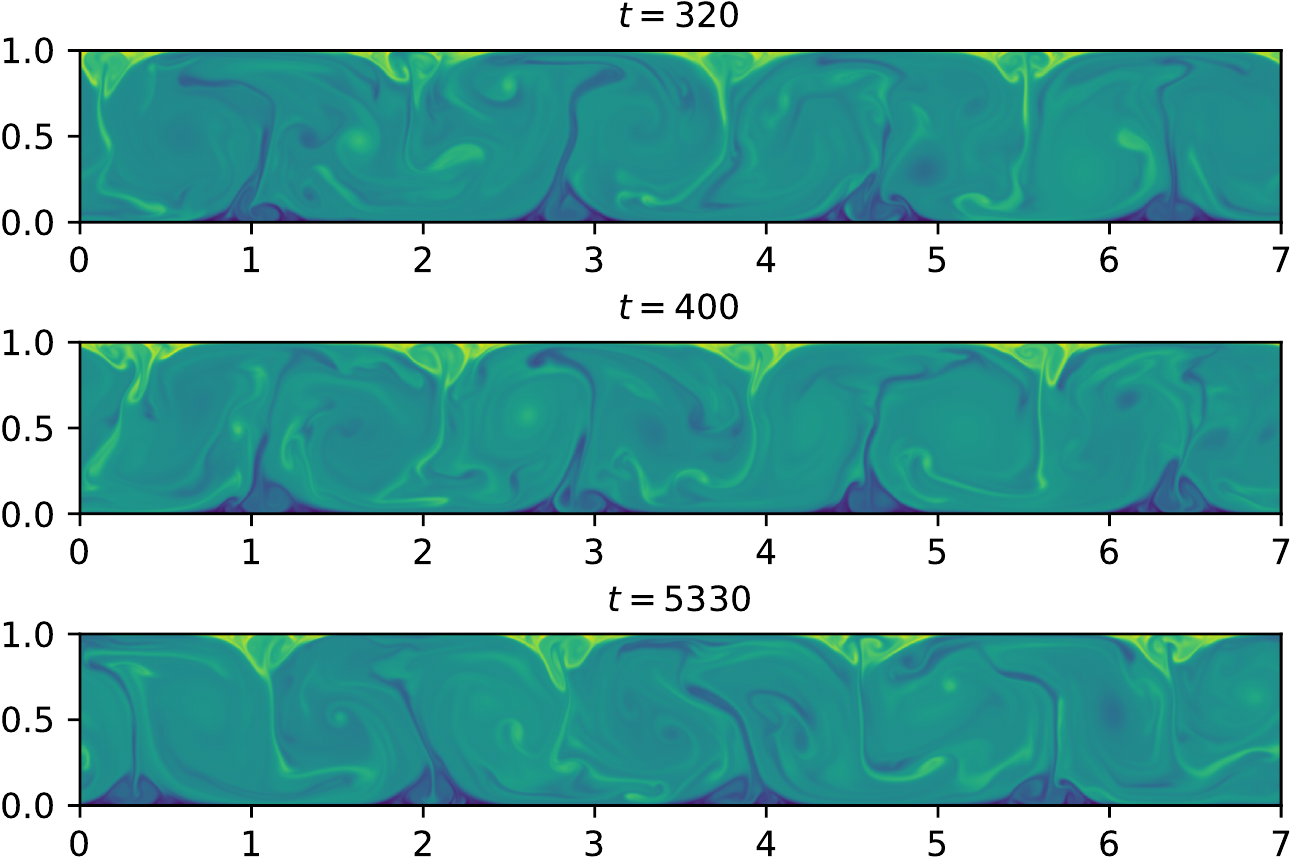}
  \caption{%
    Visualization of the temperature field for the high-resolution \ac{rbc}
    simulation, at $t = 320$ eddy turnovers, $t = 400$ eddy turnovers and
    $t = 5330$ eddy turnovers (end of the extended simulation, produced for
    validation of the training dataset).
    }
  \label{fig:RB_temp_snapshots}
\end{figure}

For an overview of the complexity of the physical patterns in the training
dataset, we show in Figure \ref{fig:RB_temp_snapshots} the temperature field at
three times. It should be noted that the training data (and, as a result, also the GAN
results) are still not completely physical, because we only treated the
two-dimensional case. Real engineering applications are all three-dimensional,
displaying additional physical phenomena (such as vortex stretching), which are
beyond the scope of the present study. Nonetheless, two-dimensional studies are
still useful, as they still share much of the physics with the three-dimensional
case (and therefore constitute a good test for the present GAN training
with constraints). A future version of this study will treat the full,
three-dimensional case.

\section{Results}
In this work, the performances of standard GAN and the statistical constrained GAN are compared by investigating three different datasets. The first dataset is generated by sampling from a 2-D Gaussian process. The second and third datasets are from the lattice Boltzmann simulations of 2-D Rayleigh-B\'enard convection with different Rayleigh number and spatial resolution. 

First, the results show that the constrained GAN better emulates the statistics of the training data by incorporating the constraint of covariance structure, indicating that the statistical constraint leads to the better converging towards the global minimum, where all statistics of training data can be captured by GANs. It should be noted that the similar performance can also be achieved by fine tuning the training process of the standard GANs. However, tuning a GAN largely relies on the experience of users and even may not be feasible for complex systems. Therefore, the constrained GAN is more stable in terms of training, making it a more practical tool for emulating complex systems.

Second, we show that the constrained GAN can achieve even higher quality of results at a significantly lower computational cost (up to 80\% reduction of computational cost in model training) compared with the unconstrained model. In effect, the addition of the statistical constraint reduces the space of allowable solutions, forcing the training procedure (here, a standard stochastic gradient descent-based method) to explore only this reduced solution space. One could argue that, even if the model was being trained by randomly selecting parameter configurations from the allowable set of solutions, the fact that the feasible set is now smaller will allow for a faster exploration on average, resulting in a reduced training time. We recognize that this is of course an experimental result on an idealized system, but these results are in line with the benefits expected from regularizing machine learning models. 

\subsection{2-D Gaussian Process}
\begin{figure}
  \centering
  \includegraphics[width=0.3\textwidth]{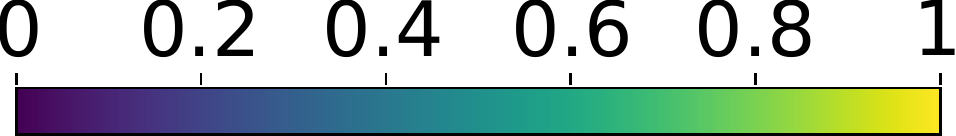}\\
  \subfloat[Training data]{\includegraphics[width=0.25\textwidth]{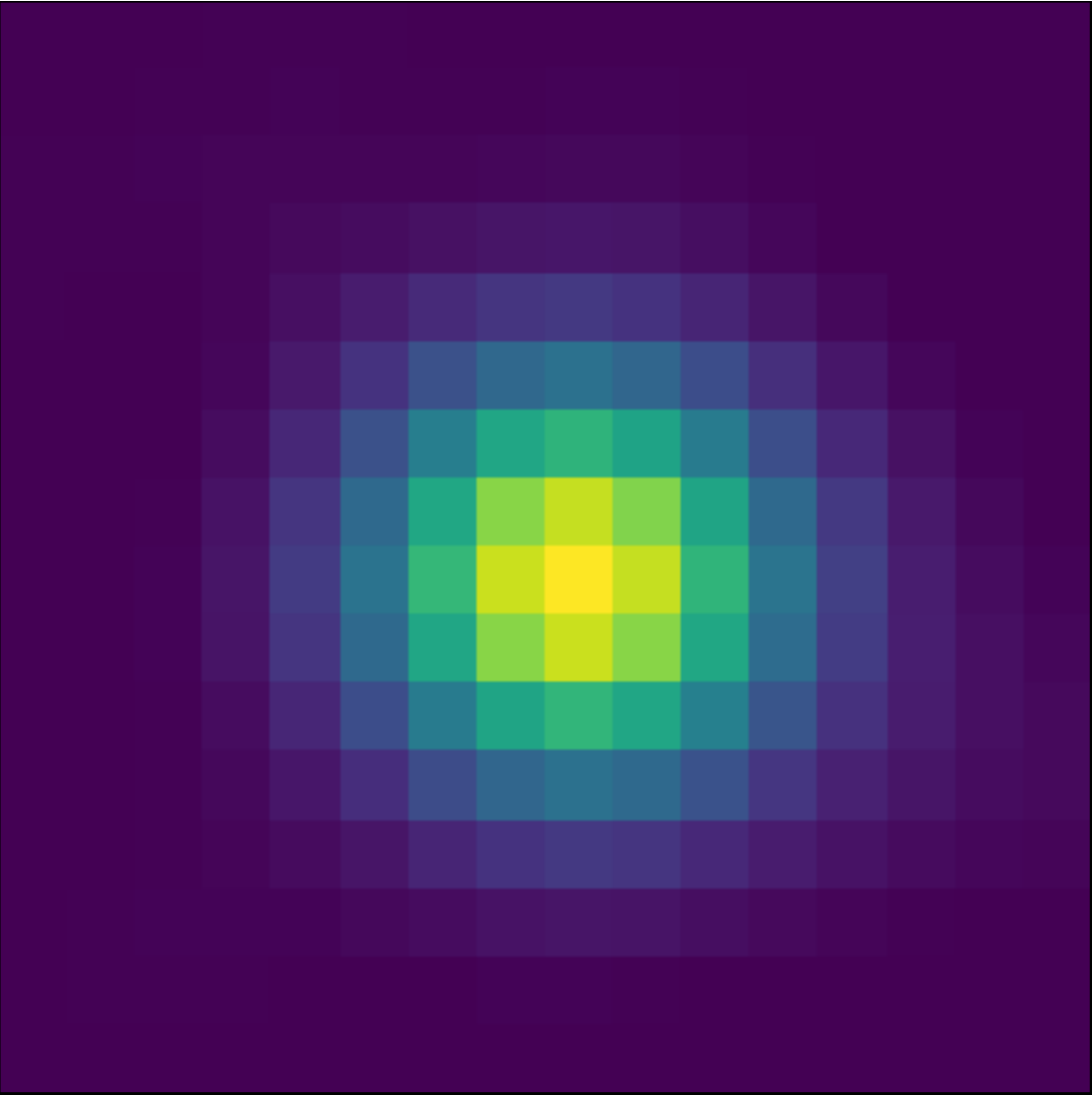}}
\hspace{0.01em}
  \subfloat[Standard]{\includegraphics[width=0.25\textwidth]{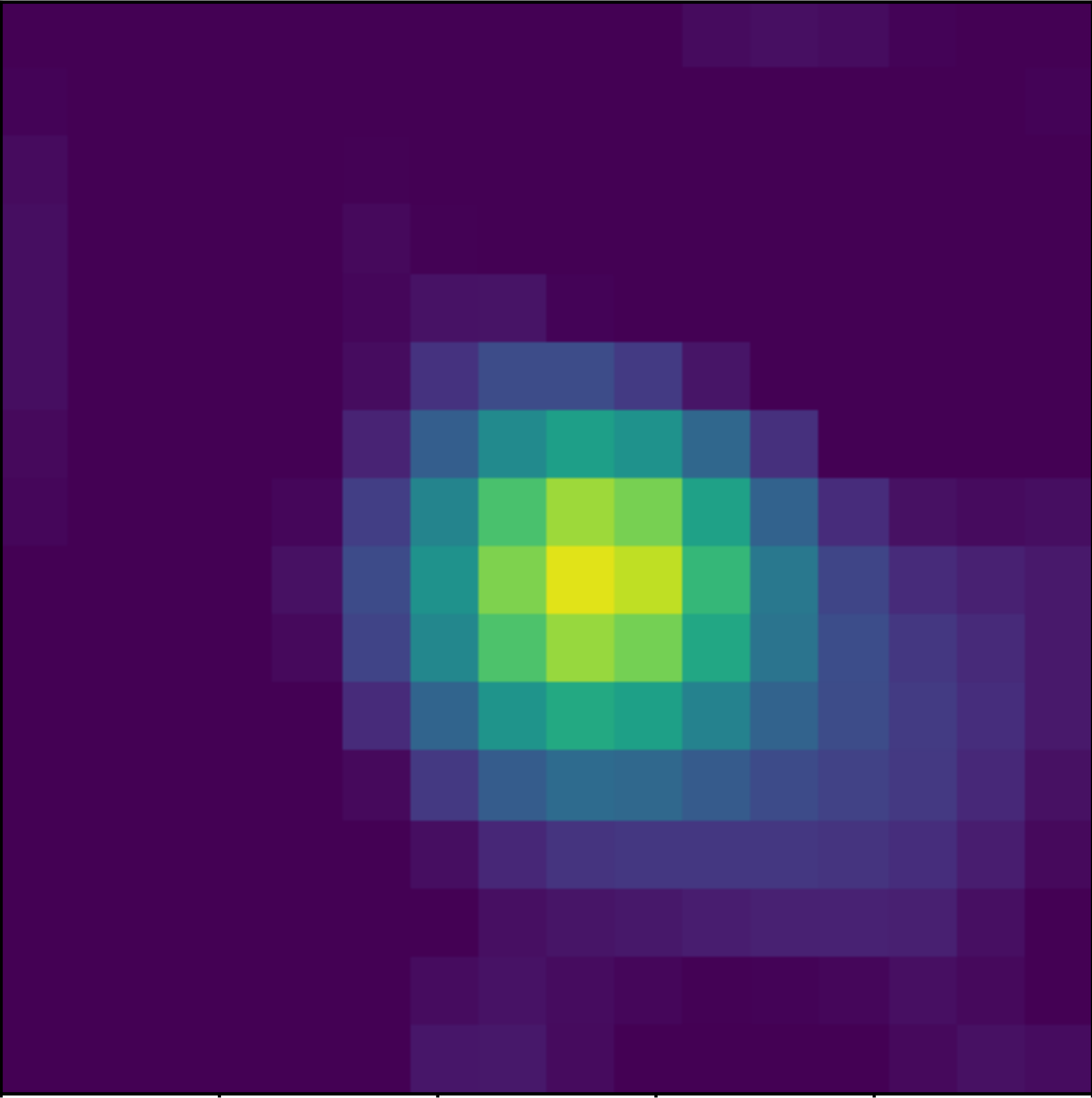}}
\hspace{0.01em}
  \subfloat[Constrained]{\includegraphics[width=0.25\textwidth]{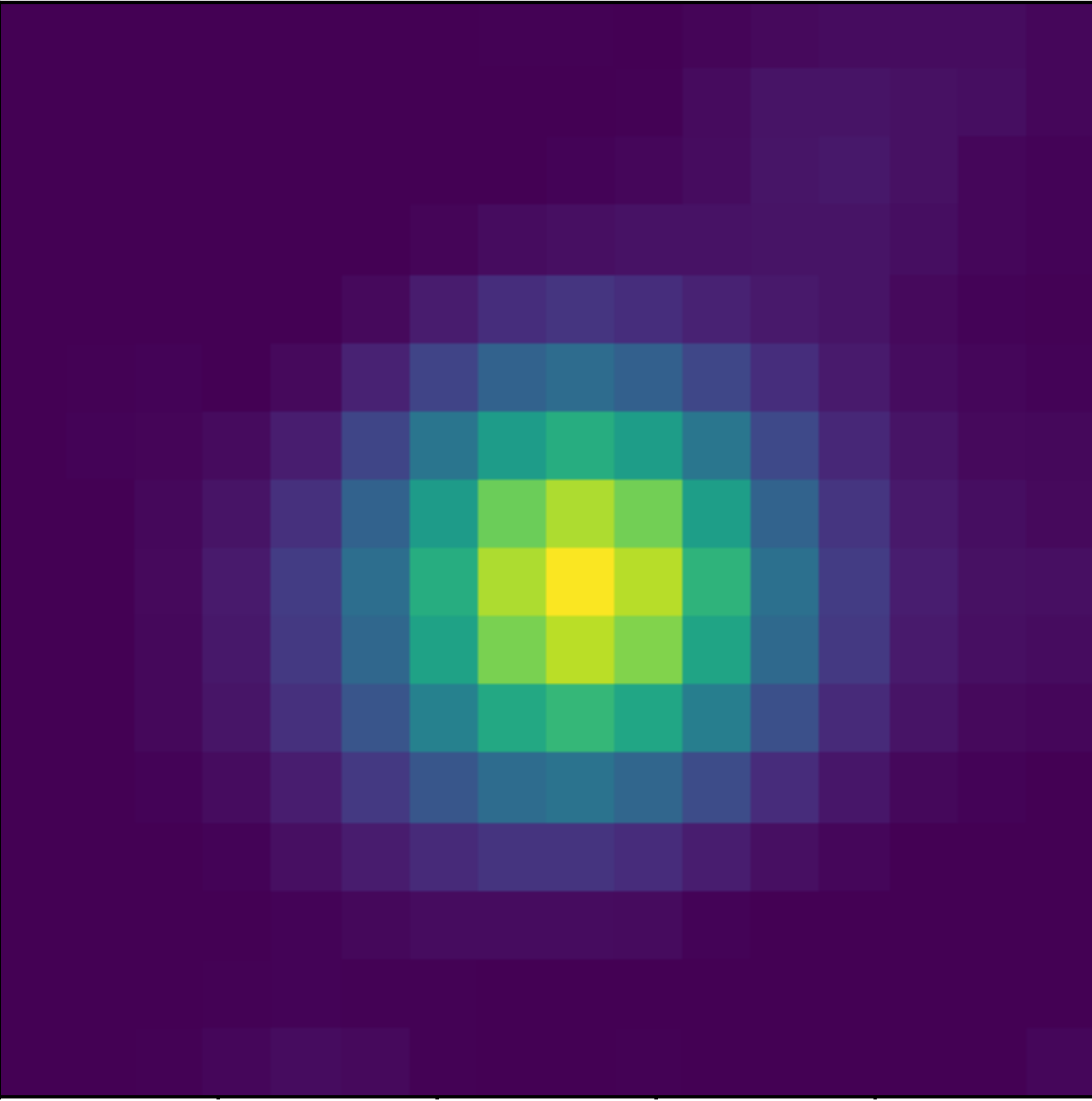}}
    \caption{Covariance with regard to the center point by using (a) training samples, (b) samples from standard GAN and (c) samples from constrained GAN. The statistical constrained GAN better captures the symmetry pattern of covariance from the training samples of the Gaussian process.}
  \label{fig:GP}
\end{figure}

We first compare the performances of standard GANs and the statistical constrained GANs by using the samples from 2-D Gaussian process as the training data. Square exponential kernel is used to specify the covariance structure $\bm{\Sigma}$ of the Gaussian process. 
\begin{equation}
\label{eq:GP_kernel}
K(\bm{x},\bm{x}^\prime)=\exp{\left(-\frac{\|\bm{x}-\bm{x}^\prime\|^2}{2\ell^2}\right)}
\end{equation}
The length scale is chosen as $\ell=0.2L$, where $L$ denotes the length of side of the square domain. The training dataset is obtained by sampling from the 2-D Gaussian process, and we acquired 10000 samples in total. These samples are provided to the discriminator of GANs with the label of 1 (true), and the objective is to use GANs to generate samples that capture the statistics of the training samples. The covariance of training data with regard to the center point is presented in Fig.~\ref{fig:GP}a, in which a symmetrical pattern can be seen and the magnitude of covariance gradually decays from the center to the side.

\begin{figure}
  \centering
  \includegraphics[width=0.6\textwidth]{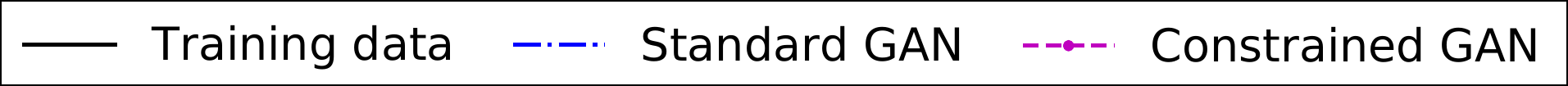}\\
  \subfloat[Diagonal 1 ]{\includegraphics[width=0.44\textwidth]{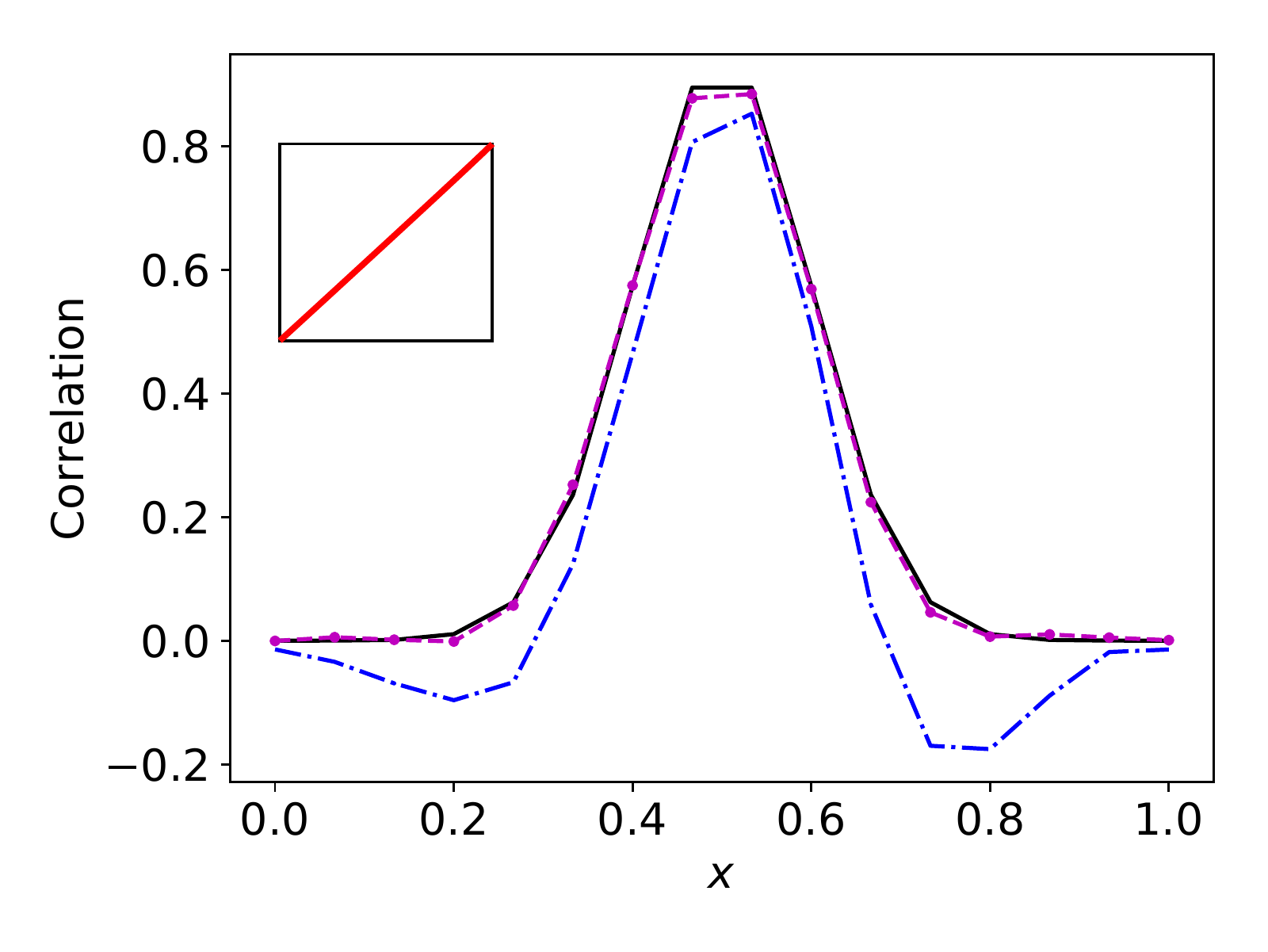}}
\hspace{0.01em}
  \subfloat[Diagonal 2]{\includegraphics[width=0.44\textwidth]{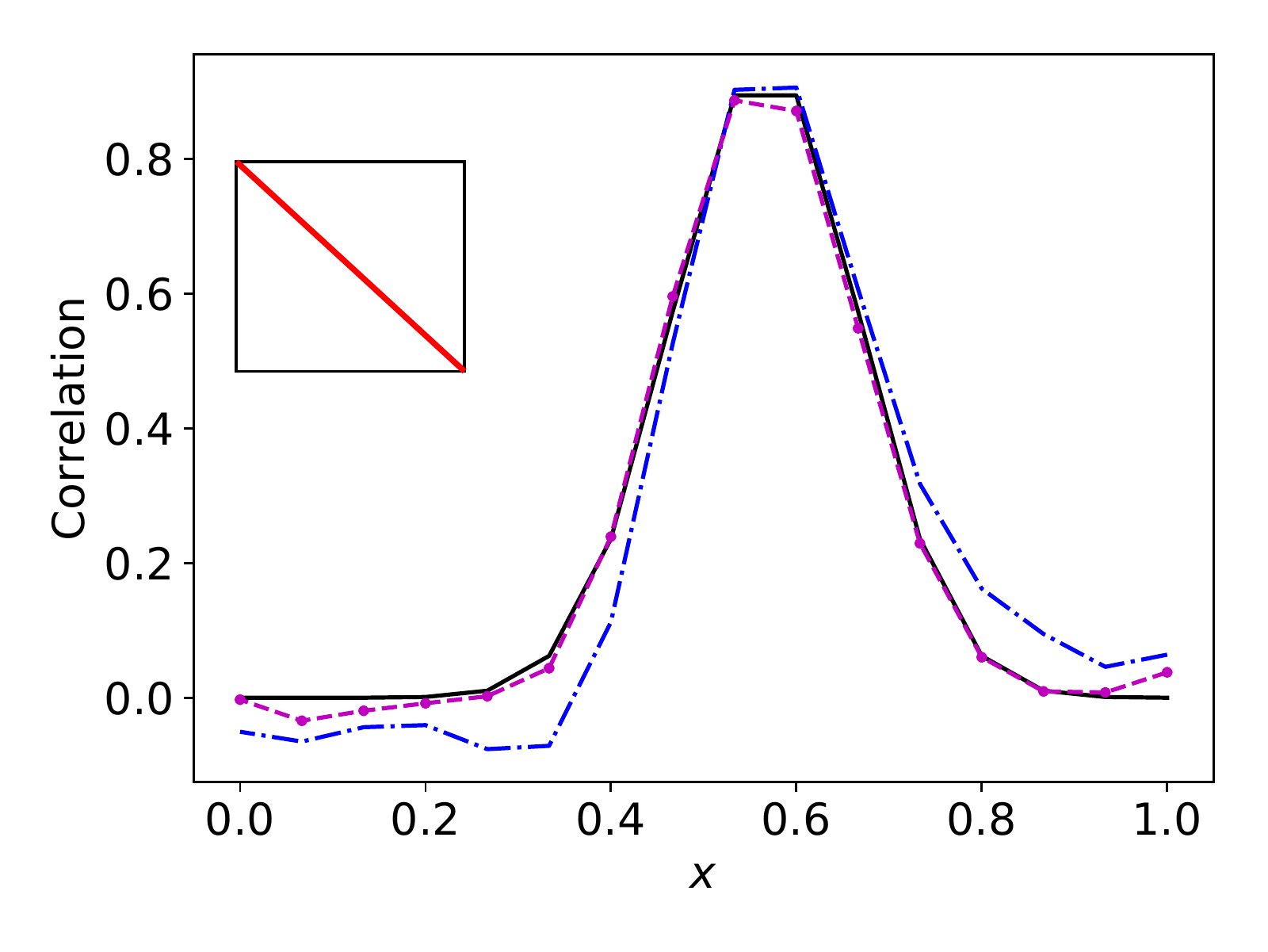}}
    \caption{Correlation profiles along (a) diagonal 1 and (b) diagonal 2. The red line within a square at the top-left part of each figure indicates the direction of the diagonal.}
  \label{fig:GP-diag}
\end{figure}

The comparison of the covariance of the generated samples from GANs demonstrates the superior performance of the statistical constrained GAN as shown in Fig.~\ref{fig:GP}. Specifically, it can be seen in Fig.~\ref{fig:GP}b that the estimated covariance of the generated samples from the standard GAN shows a noticeable asymmetrical pattern. More quantitative comparison of correlation profiles along diagonals in Fig.~\ref{fig:GP-diag} also confirms that the results from constrained GAN better capture the symmetrical pattern of the correlation field from training data. It indicates that the standard GAN converges to a local minimum and the statistics of the training data is not truthfully reproduced. On the contrary, the covariance of the generated samples from the statistical constrained GAN shown in Fig.~\ref{fig:GP}c demonstrates a better agreement with the training data. It indicates that the statistical constraint guide GANs to better converge toward global minimum.

\begin{figure}[!htbp]
  \centering
  \includegraphics[width=0.5\textwidth]{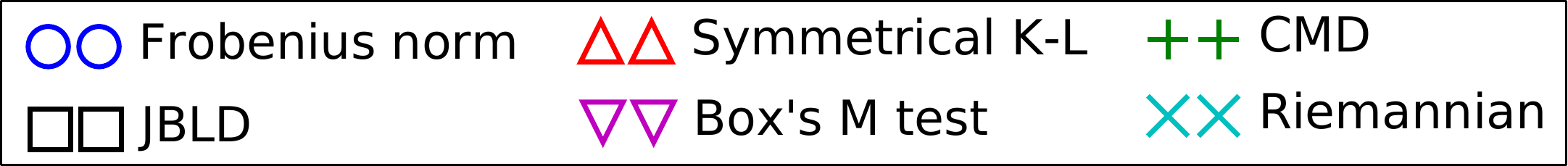}\\
  \subfloat[Standard GAN]{\includegraphics[width=0.45\textwidth]{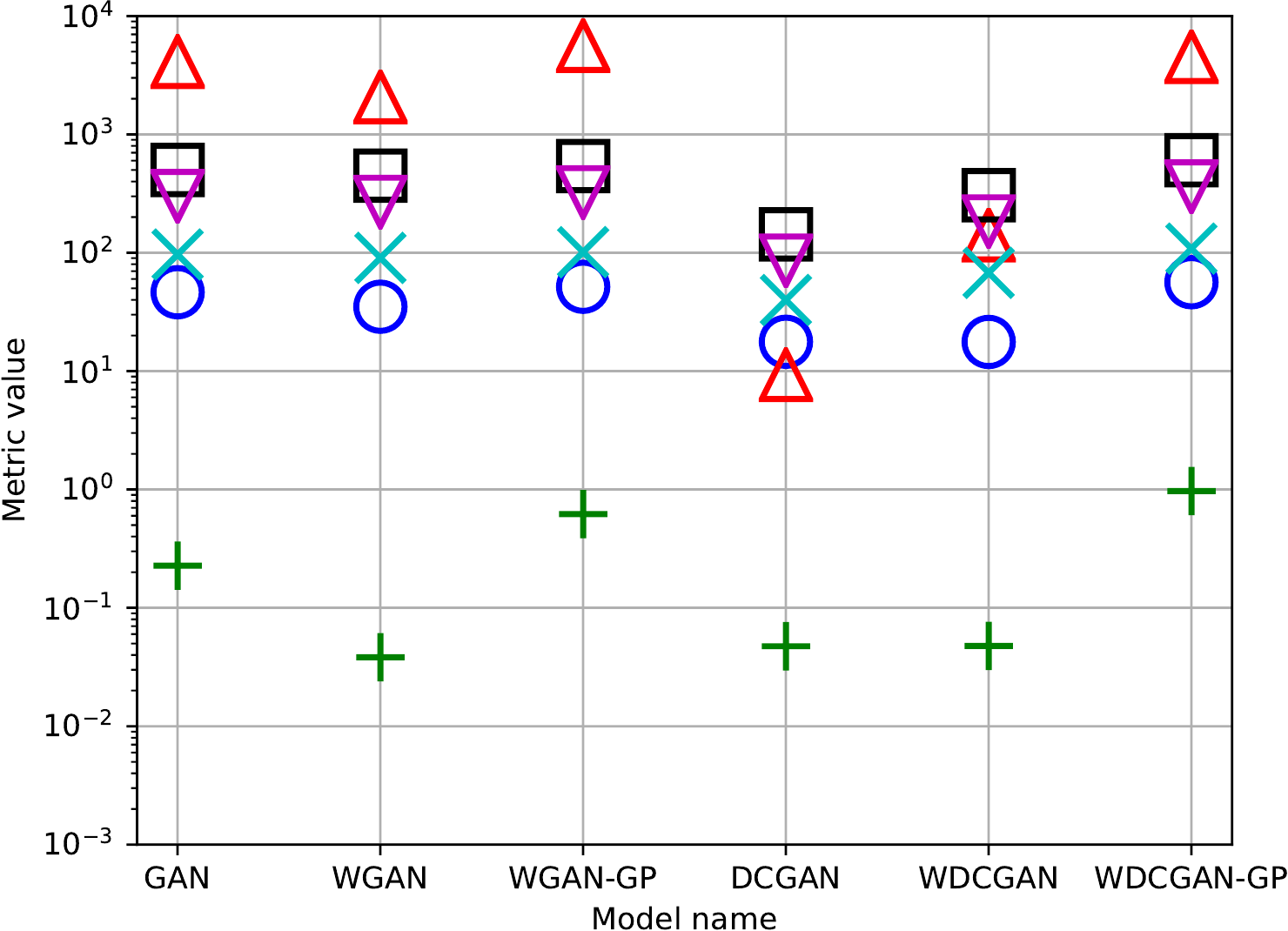}}
  \subfloat[Constrained GAN]{\includegraphics[width=0.45\textwidth]{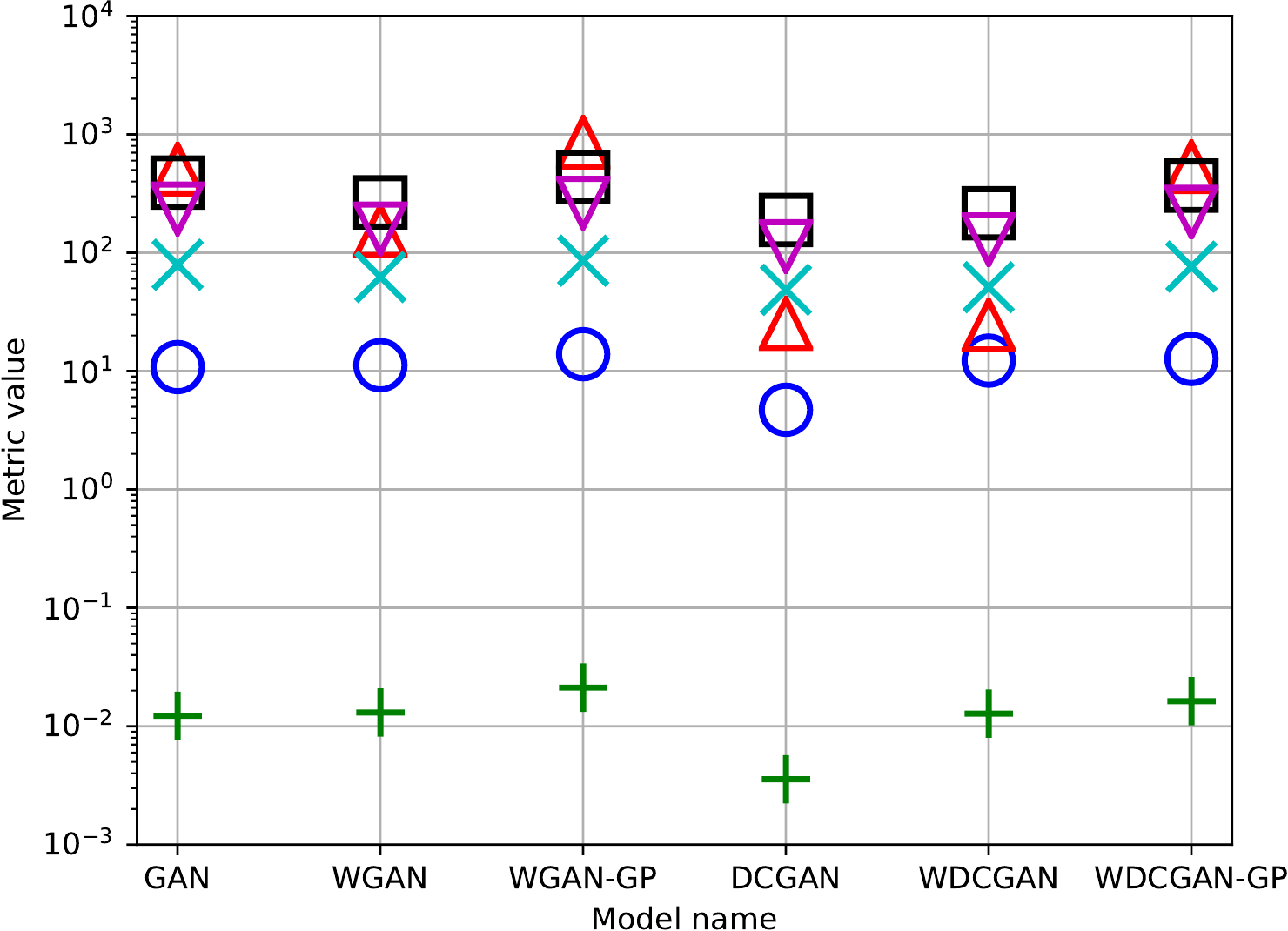}}
    \caption{Summary of the distance metrics of the covariance matrix from the generated samples of (a) standard GAN and (b) constrained GAN. Specifically, WGAN denotes \textit{Wasserstein} GAN~\cite{arjovsky2017wasserstein}, and GP denotes gradient penalty as proposed in~\cite{gulrajani2017improved}.}
  \label{fig:GP-dist-metrics}
\end{figure}

The main purpose of introducing the statistical constraint can be interpreted in two ways. First, the constraint term serves as a regularization and reduces the loss function to a lower dimensional manifold, in which the convergence to global minimum would be easier to achieve. Second, the constraint term contributes a non-zero gradient when the training process get into local minima and thus the optimization is unlikely to stay at a local minimum. According to these two purposes, it is not necessary to define a precise metric of the difference between the distributions of the training data and the generated samples. Instead, an approximate distance metric may work well enough as long as it vanishes when two distributions are identical to each other. As shown in Eq.~\ref{eq:loss_func}, the distance between two covariance matrices is defined in Euclidean space by using Frobenius norm. The main reason of using Frobenius norm is the relatively low computational cost, compared with other distance metrics, e.g., the K--L divergence and the Riemannian distance. After obtaining the generated samples, we evaluated different distance metrics to quantify the difference between the statistics of the training data and the generated samples. It can be seen in Fig.~\ref{fig:GP-dist-metrics} that the statistical constraint term generally leads to smaller distance metrics when being implemented into different GANs architectures. It confirms that the constrained GAN indeed better converges toward the global minimum, instead of merely reducing the specific distance metric adopted in the training.

\begin{figure}[!htbp]
  \centering
  \includegraphics[width=0.35\textwidth]{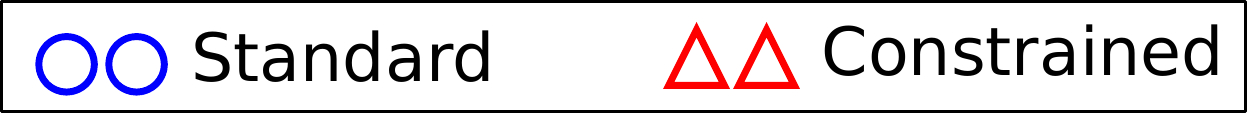}\\
  \subfloat[Skewness]{\includegraphics[width=0.44\textwidth]{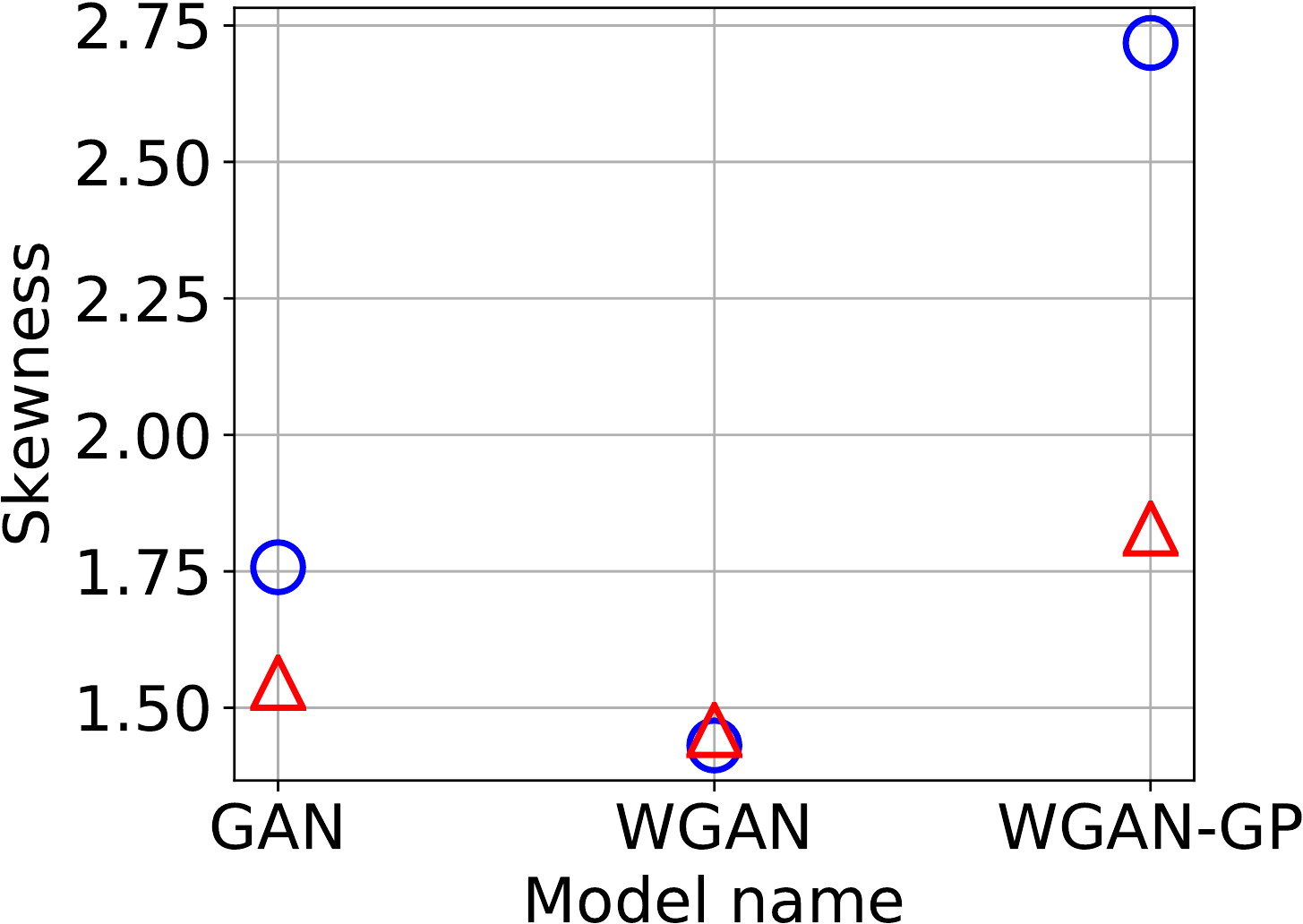}}
  \subfloat[Kurtosis]{\includegraphics[width=0.44\textwidth]{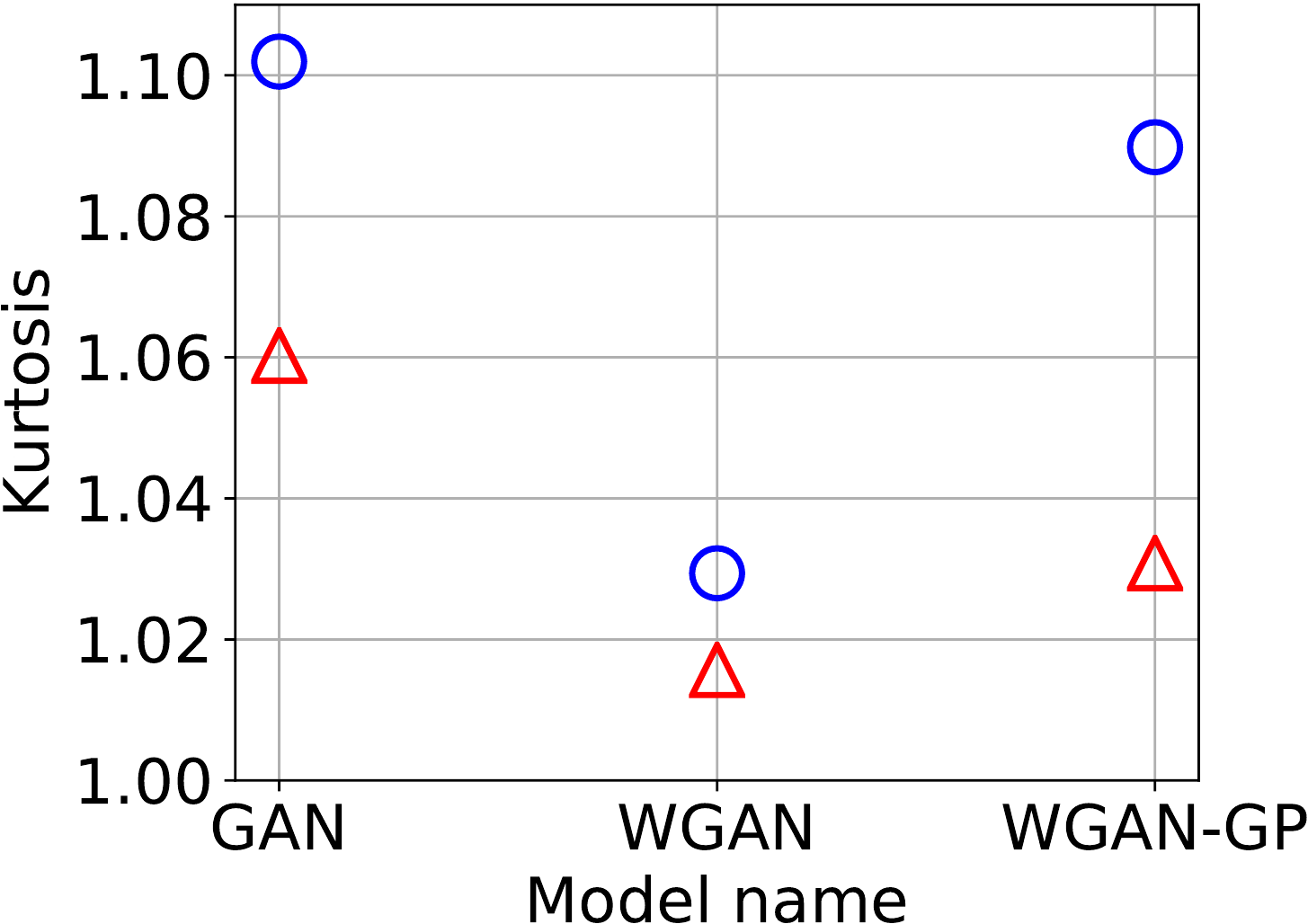}}
    \caption{High-order statistics of (a) skewness and (b) kurtosis provided by different types of GANs. The physics-informed constraint leads to improvement of skewness and kurtosis for most types of GANs.}
  \label{fig:GP-high-order}
\end{figure}

We also investigated high-order statistics to illustrate the advantage of the statistical constrained GAN. The skewness and the kurtosis of the generated samples in Fig.~\ref{fig:GP-high-order} show that the high-order statistics of the samples have a better agreement with the training data when using the statistical constrained GAN, indicating that high-order statistics are also better captured by introducing the covariance constraint. It should be noted that both the skewness and the kurtosis in Fig.~\ref{fig:GP-high-order} are normalized by the benchmark values of the multivariate Gaussian distribution, and thus the ideal values should be one. As shown in Fig.~\ref{fig:GP-high-order}a, the larger skewness of generated samples from standard GAN indicates that the sample distribution is more asymmetrical, which has been confirmed by the visualization of covariance in Fig.~\ref{fig:GP}.

\subsection{Rayleigh-B\'enard convection at a low Rayleigh number}
\begin{figure}[!htbp]
  \centering
  \includegraphics[width=0.3\textwidth]{field-legend}\\
  \subfloat[Training data]{\includegraphics[width=0.3\textwidth]{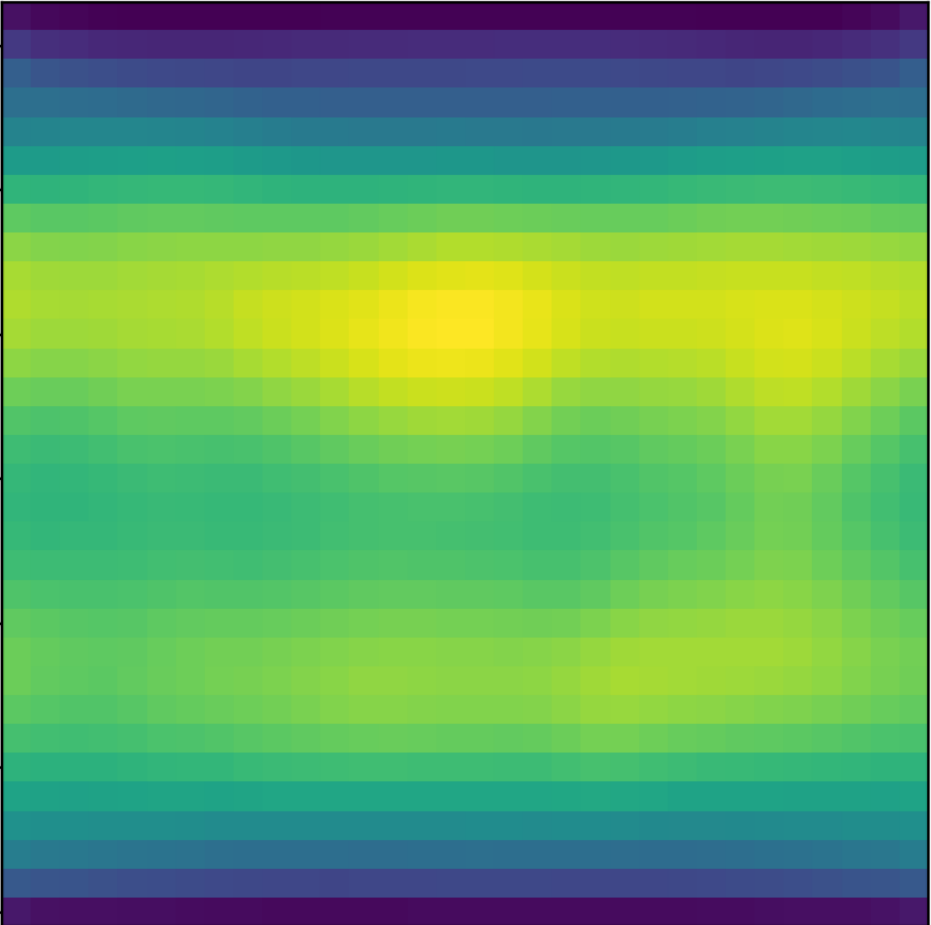}}\hspace{0.2em}
  \subfloat[Standard GAN]{\includegraphics[width=0.3\textwidth]{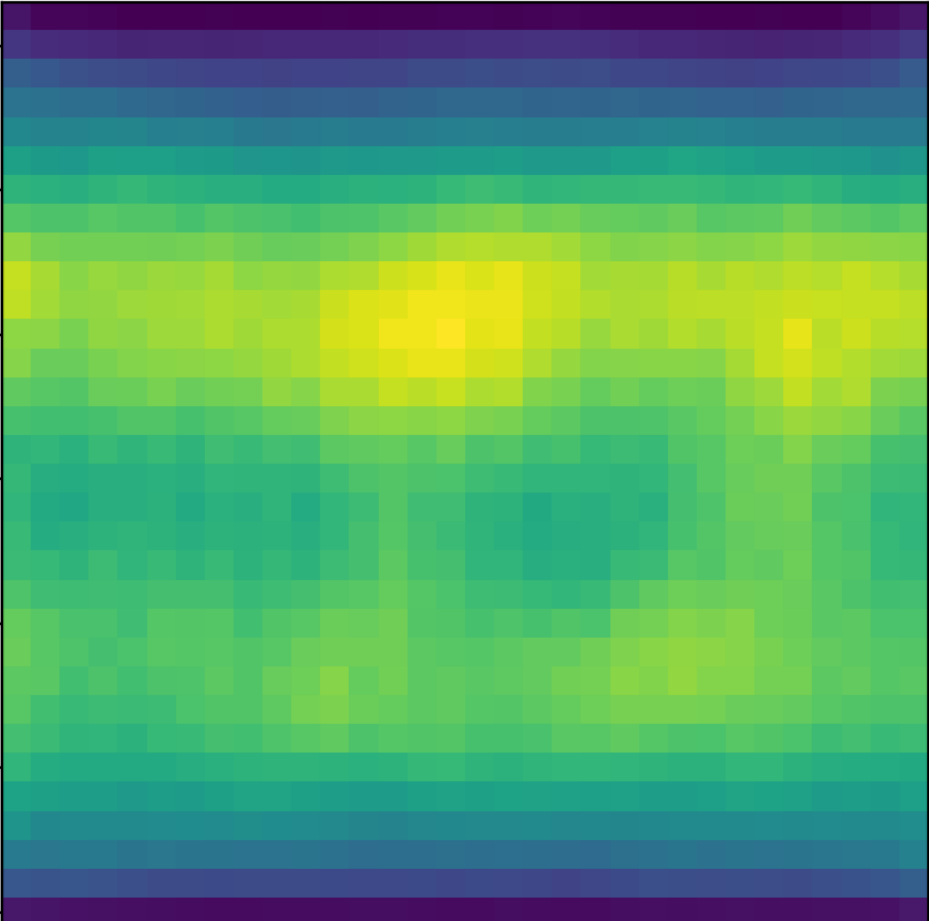}}
\hspace{0.2em}
  \subfloat[Constrained GAN]{\includegraphics[width=0.3\textwidth]{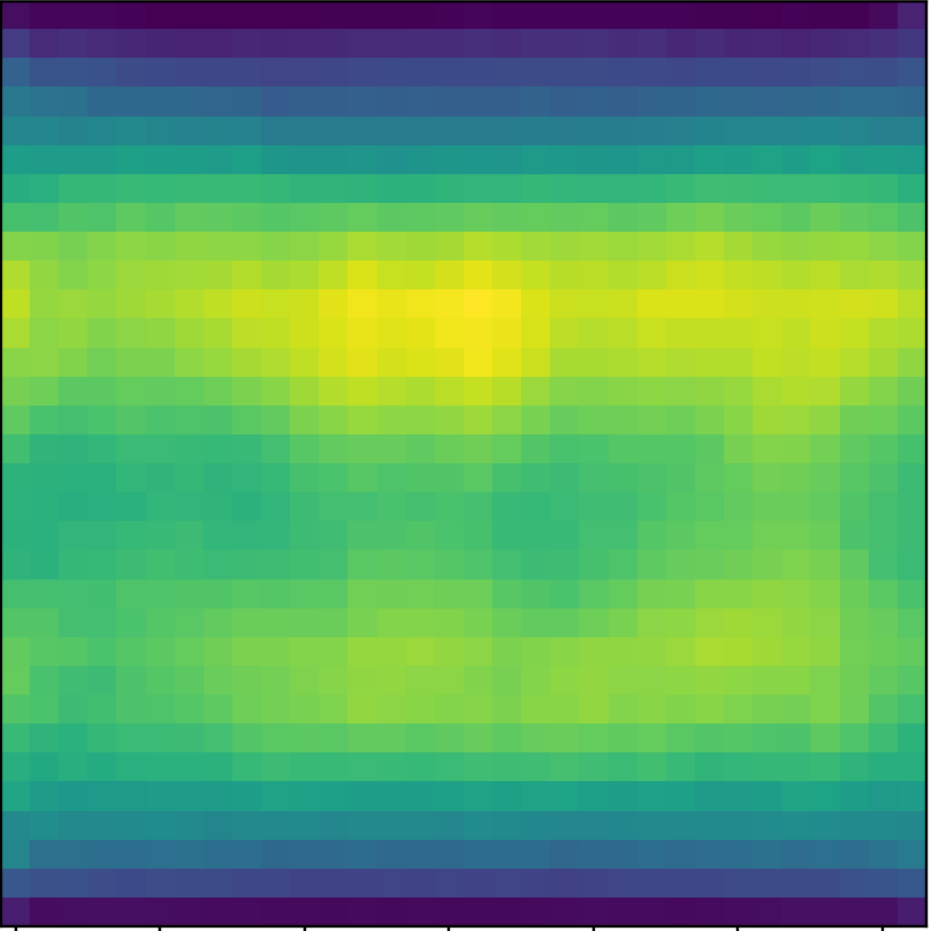}}
    \caption{Spatial distribution of turbulent kinetic energy of 2-D Rayleigh-B\'enard convection from (a) training samples, (b) generated samples from GAN and (c) generated samples from physics-informed GAN. The learning rate of the generator is set as 0.002.}
  \label{fig:RBC-tke-comp}
\end{figure}

We also studied the performance of the statistical constrained GAN by using training data from the simulations of Rayleigh-B\'enard convection. The training data is obtained from a further simplified RBC system compared to the one described in Section~\ref{sec:LBM}, in order to first investigate a less chaotic RBC system (Rayleigh number $Ra=10000$ and Prandtl number $Pr=10$). Specifically, the simulation is performed in a square domain. The top wall is at a low temperature and the bottom wall is at a high temperature. The two side walls have the periodic boundary condition. The training data corresponds to 10000 snapshots of the instantaneous flow field. The turbulent kinetic energy (TKE) of the training data is presented in Fig.~\ref{fig:RBC-tke-comp}. It can be seen that the TKE is relatively low near both the top and the bottom walls. In addition, higher TKE can be observed within two horizontal stripe-shape regions both above and below the center region. By carefully tuning the learning rate, it can be seen in Fig.~\ref{fig:RBC-tke-comp} that the TKE of generated samples shows a good agreement with the training data by using either the standard GAN or the statistical constrained GAN. It is because the standard GAN is capable of preserving all the statistics of the training data if global minimum is achieved for the training process. Therefore, with a proper choice of training parameters, the performances of the standard GAN and the statistical constrained GAN should be comparable with each other, which has been confirmed in Fig.~\ref{fig:RBC-tke-comp}.

\begin{figure}[!htbp]
  \centering
  \includegraphics[width=0.3\textwidth]{field-legend}\\
  \subfloat[Training data]{\includegraphics[width=0.3\textwidth]{tke}}\hspace{0.2em}
  \subfloat[Standard GAN]{\includegraphics[width=0.3\textwidth]{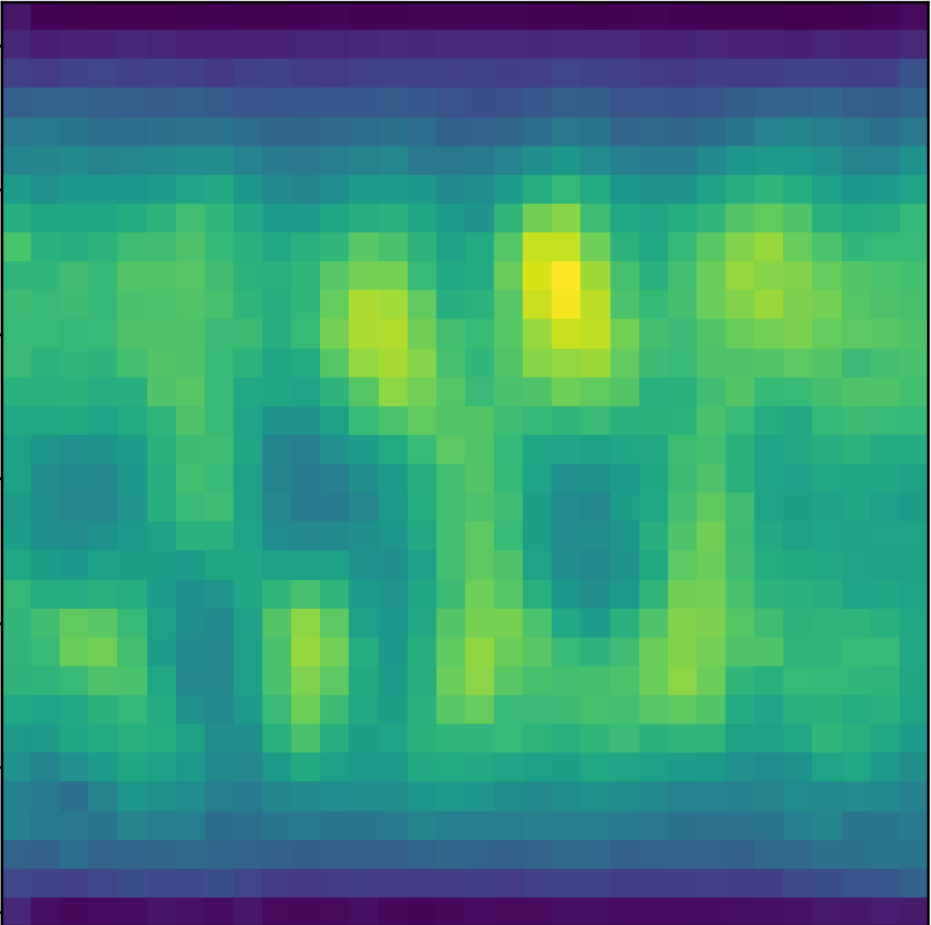}}
\hspace{0.2em}
  \subfloat[Constrained GAN]{\includegraphics[width=0.3\textwidth]{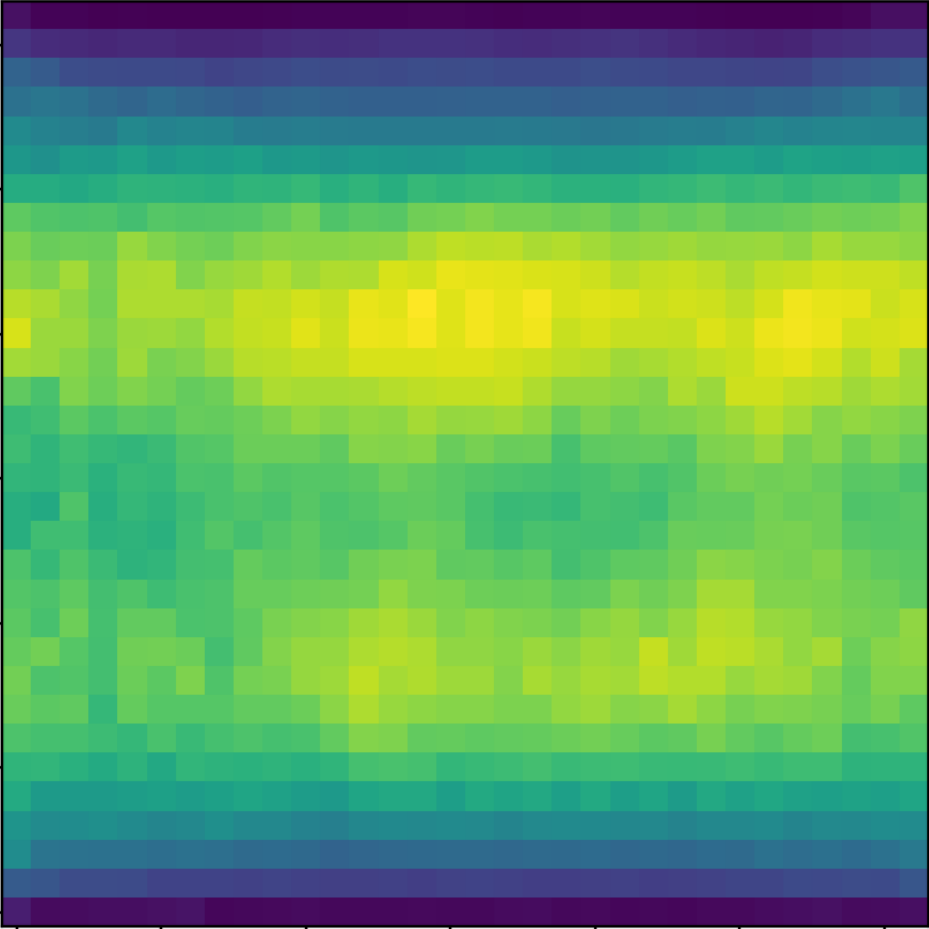}}
    \caption{Spatial distribution of turbulent kinetic energy of 2-D Rayleigh-B\'enard convection from (a) training samples, (b) generated samples from GAN and (c) generated samples from physics-informed GAN. The learning rate of the generator is set as 0.02.}
  \label{fig:RBC-tke-comp2}
\end{figure}

However, the optimal training is usually difficult to be achieved, especially for training a GAN to emulate more complex physical systems. In order to illustrate the stability of training GANs, we adopt another learning rate $lr=0.02$ and present the TKE of generated samples in Fig.~\ref{fig:RBC-tke-comp2}. It can be seen in Fig.~\ref{fig:RBC-tke-comp2}b that the result of standard GAN changes significantly, showing noticeable difference from the training data. Although the higher TKE regions generated by the standard GAN still locates above and below the horizontal center region, the regions with high TKE become less continuous and do not capture the pattern in Fig.~\ref{fig:RBC-tke-comp}. On the contrary, the results of statistical constrained GAN in Fig.~\ref{fig:RBC-tke-comp2}b demonstrates less changes. It should be noted that the more noisy result in Fig.~\ref{fig:RBC-tke-comp2}c compared with the one in Fig.~\ref{fig:RBC-tke-comp}c is mainly due to the larger learning rate, which introduces more noises during the training. According to the comparison shown in Fig.~\ref{fig:RBC-tke-comp2}, the statistical constrained GAN is more stable with regard to the change of training parameters.

\begin{figure}[!htbp]
  \centering
  \includegraphics[width=0.6\textwidth]{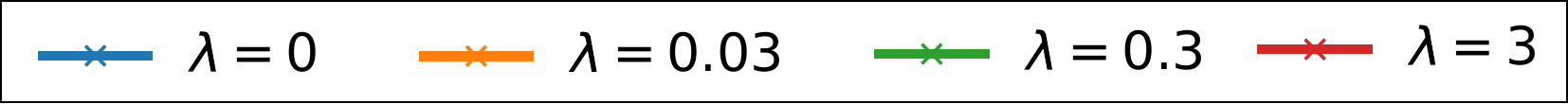}\\
  \subfloat[$lr=0.001$]{\includegraphics[width=0.49\textwidth]{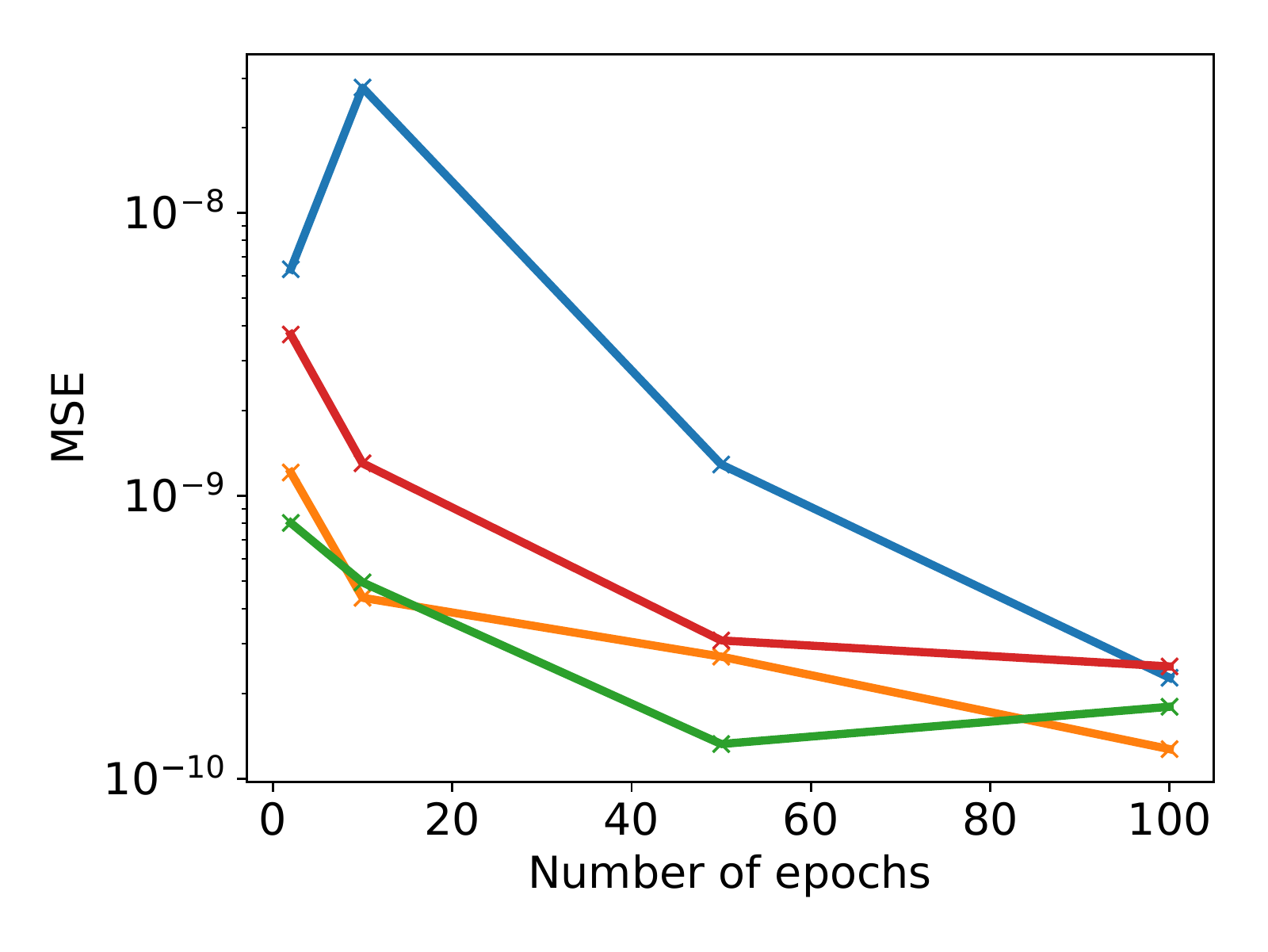}}
\hspace{0.2em}
  \subfloat[$lr=1$]{\includegraphics[width=0.49\textwidth]{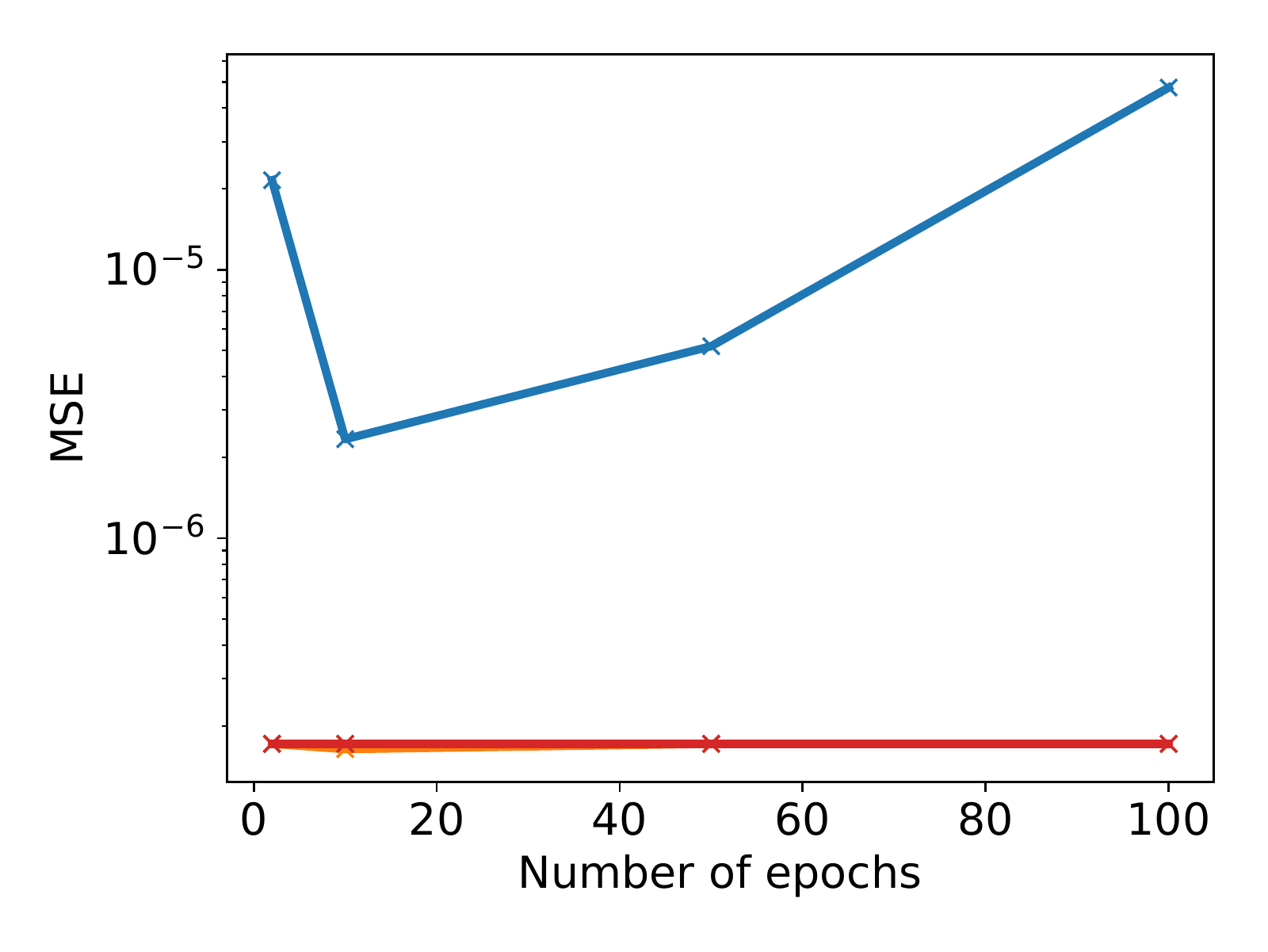}}
\hspace{0.2em}
    \caption{Mean squared errors of the predicted turbulent kinetic energy of 2-D Rayleigh-B\'enard convection with learning rates (a) $lr=0.001$, (b) $lr=1$. Four different coefficients $\lambda$ of the constraint term are investigated. It should be noted that $\lambda=0$ corresponds to standard GANs.}
  \label{fig:RBC-lr-comp}

\end{figure}

In order to illustrate the more stable training by using the statistical constrained GAN, we studied the training of GANs with different learning rate and penalty coefficient of the statistical constraint term. The comparison in Fig.~\ref{fig:RBC-lr-comp} demonstrates that better performance can be achieved by introducing the statistical constraint term. Specifically, the mean squared error (MSE) of the TKE field against the number of epochs is presented. It can be seen in Fig.~\ref{fig:RBC-lr-comp}a that the performances are comparable at 100 training epochs by using different values of penalty coefficient $\lambda$. However, with the same number of epochs, the mean squared errors of the statistical constrained GAN ($\lambda>0$) are generally smaller than the standard GAN ($\lambda=0$). By using a larger learning rate as shown in Fig.~\ref{fig:RBC-lr-comp}b, it can be seen that the training of standard GAN ($\lambda=0$) is not able to converge and the MSE is noticeably larger than the results from statistical constrained GANs.

\begin{figure}[!htbp]
  \centering
  \includegraphics[width=0.4\textwidth]{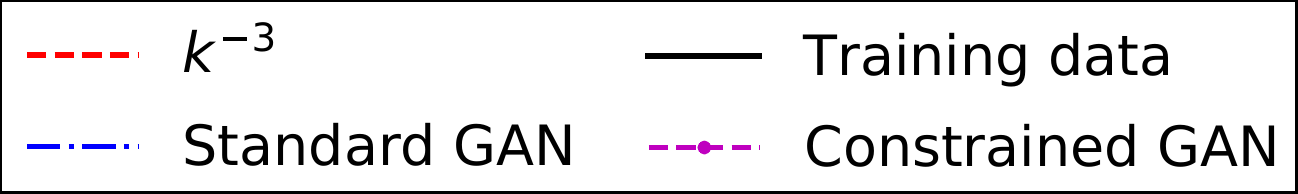}\\
  \subfloat[Learning rate=0.002]{\includegraphics[width=0.44\textwidth]{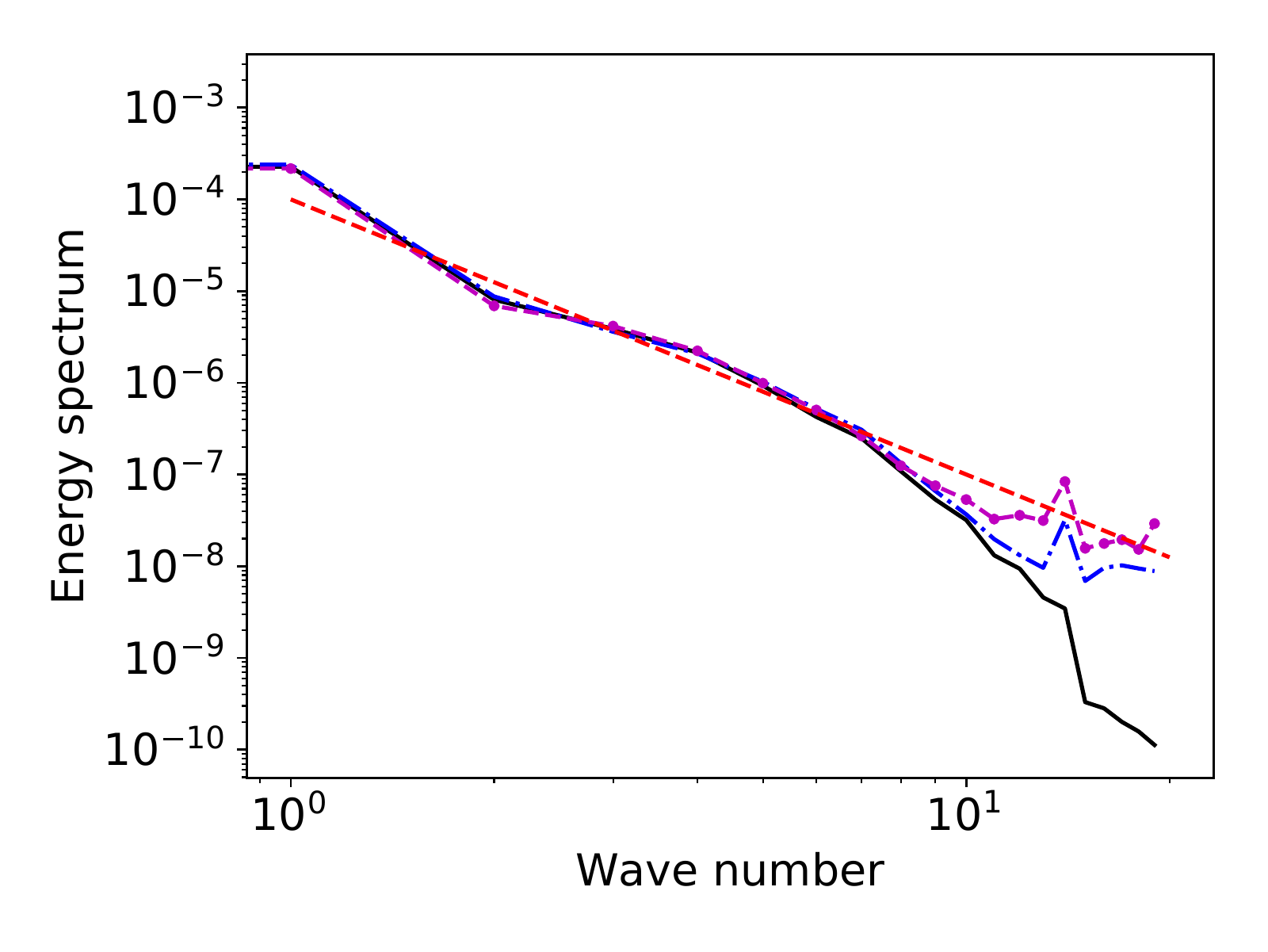}}
  \hspace{0.2em}
  \subfloat[Learning rate=0.02]{\includegraphics[width=0.44\textwidth]{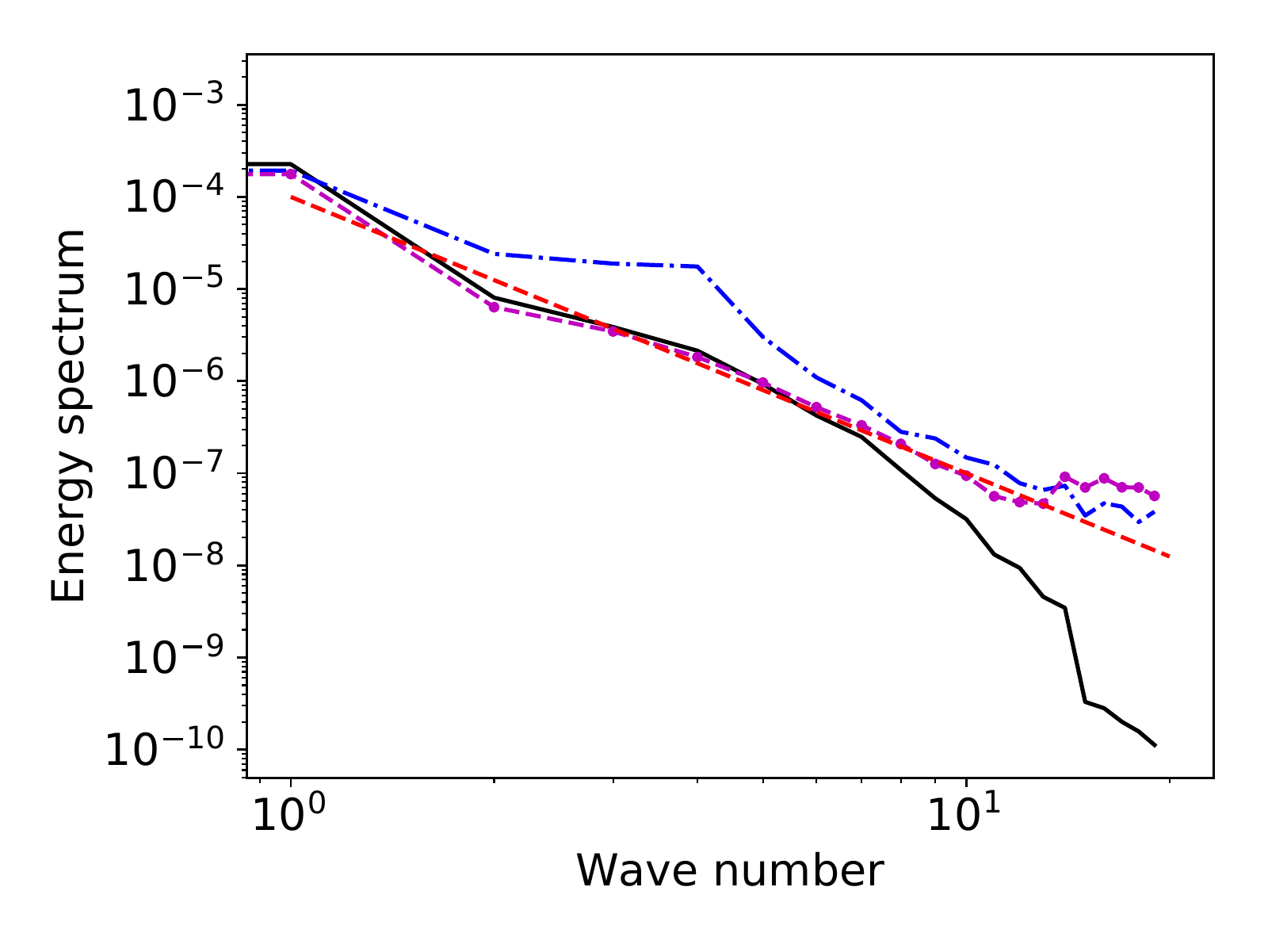}}
    \caption{Spectrum comparison of turbulent kinetic energy of 2-D Rayleigh-B\'enard convection by using different learning rates.}
  \label{fig:RBC-spectrum}
\end{figure}

The comparison of spectrum of TKE in Fig.~\ref{fig:RBC-spectrum} also confirms that the statistical constrained GAN generates samples that better capture the pattern of the training data. Although such an improvement is marginal with learning rate $lr=0.002$, it can still be seen in Fig.~\ref{fig:RBC-spectrum}a that the spectrum from the statistical constrained GAN has a better agreement with the training data for relatively small wave numbers, e.g., between 2 to 5. At the region with large wave number, both the standard GAN and the statistical constrained GAN provides energy overpredict the level of energy spectrum, which is mainly because of the small scale noises in the generated data as shown in Fig.~\ref{fig:RBC-tke-comp}. If a less proper learning rate is chosen, it can be seen in Fig.~\ref{fig:RBC-spectrum}b that the energy spectrum provided by the statistical constrained GAN shows a much better performance than the result of standard GAN, whose energy spectrum demonstrates disagreement with the training data across the whole range of wave numbers.

\subsection{Rayleigh-B\'enard convection at a high Rayleigh number}
\begin{figure}[!htbp]
  \centering
  \includegraphics[width=0.5\textwidth]{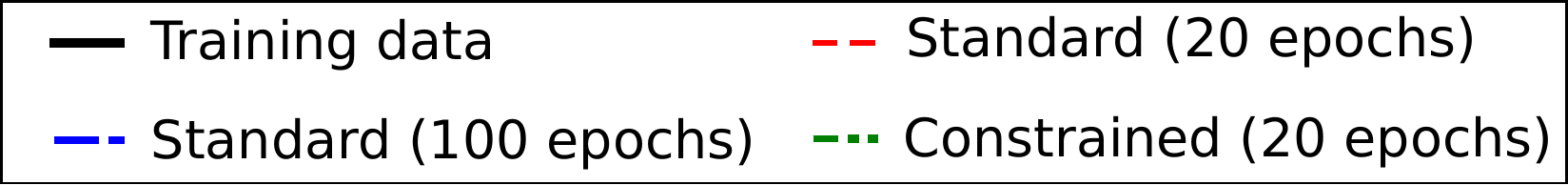}
  \subfloat[Horizontal velocity $U_x$]{\includegraphics[width=0.49\textwidth]{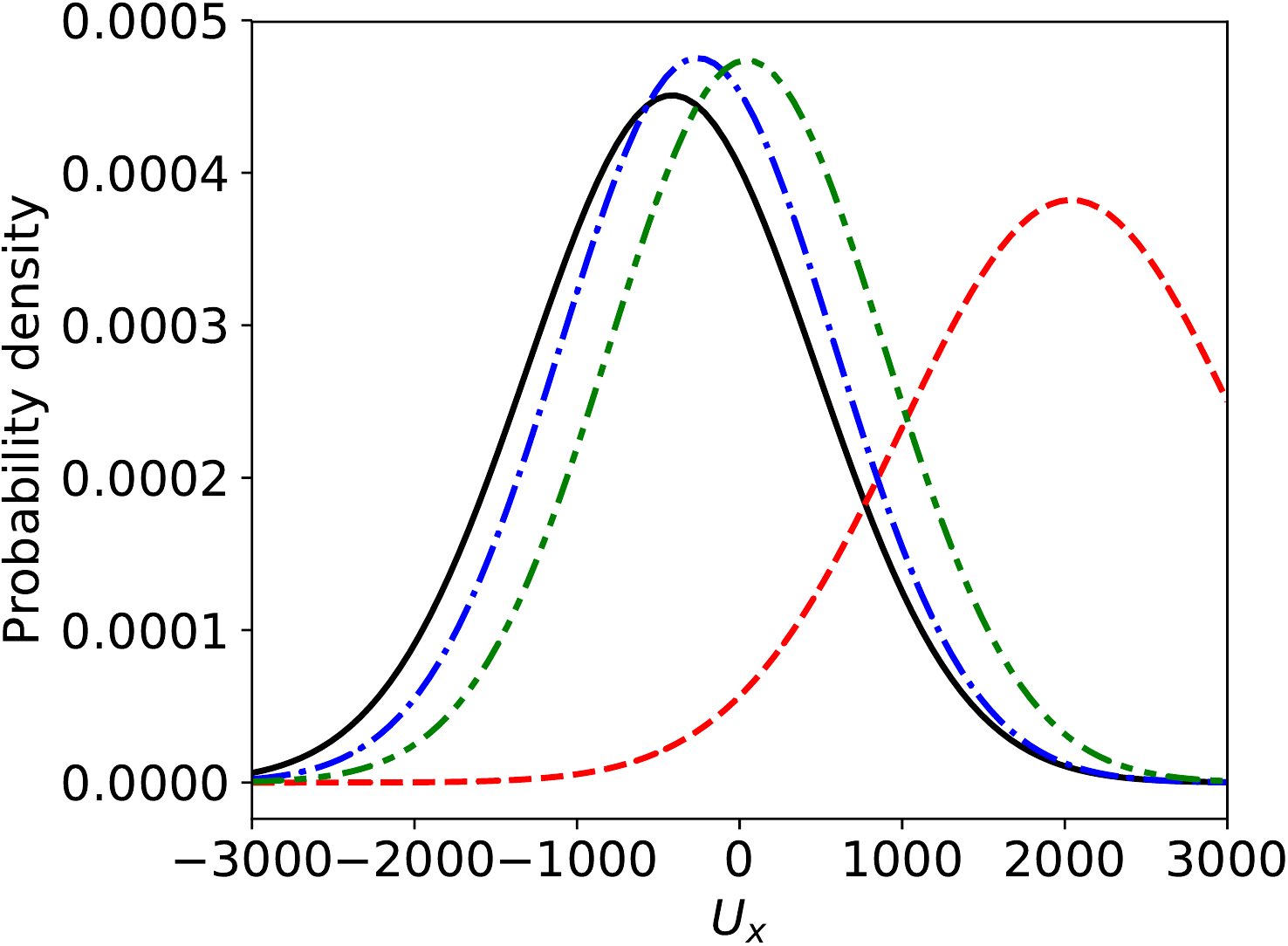}}
  \subfloat[Vertical velocity $U_y$]{\includegraphics[width=0.49\textwidth]{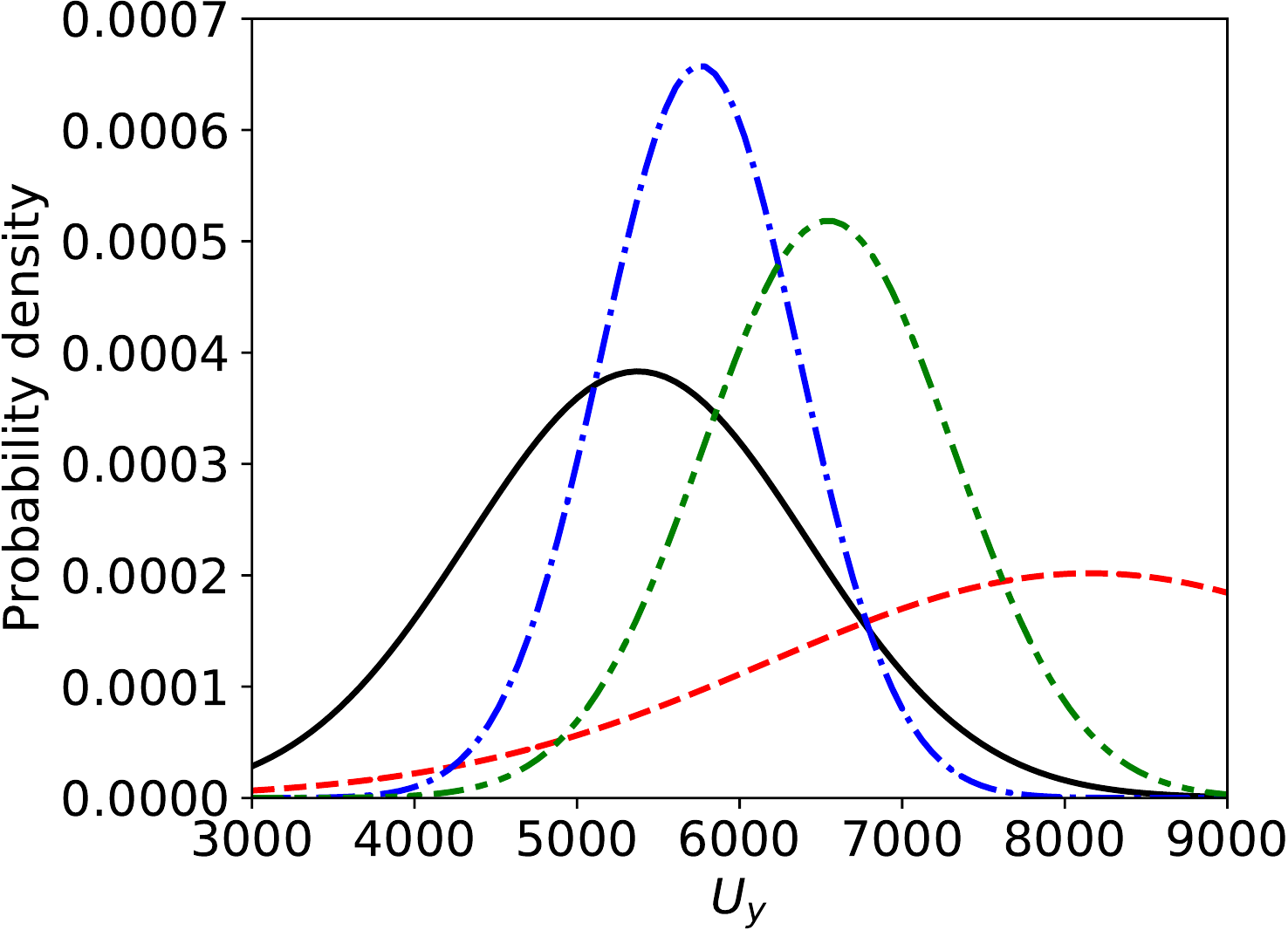}}
    \caption{The comparison of (a) horizontal velocity and (b) vertical velocity distribution of high Rayleigh number 2-D Rayleigh-B\'enard convection at the \textbf{center} point of the training domain.}
  \label{fig:RBC-U-center}
\end{figure}

\begin{figure}[!htbp]
  \centering
  \includegraphics[width=0.5\textwidth]{pdf-legends}
  \subfloat[Horizontal velocity $U_x$]{\includegraphics[width=0.49\textwidth]{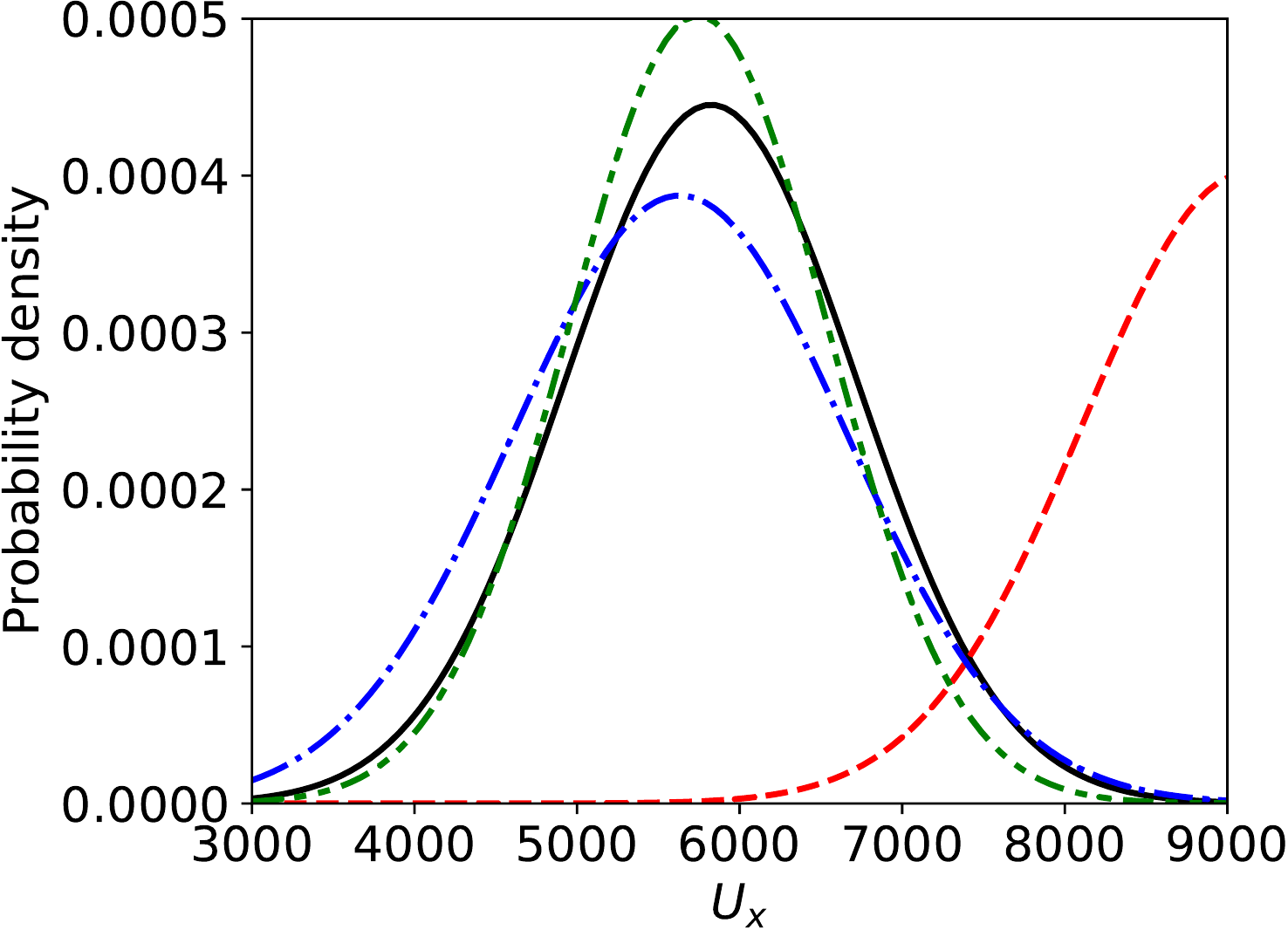}}
  \subfloat[Vertical velocity $U_y$]{\includegraphics[width=0.49\textwidth]{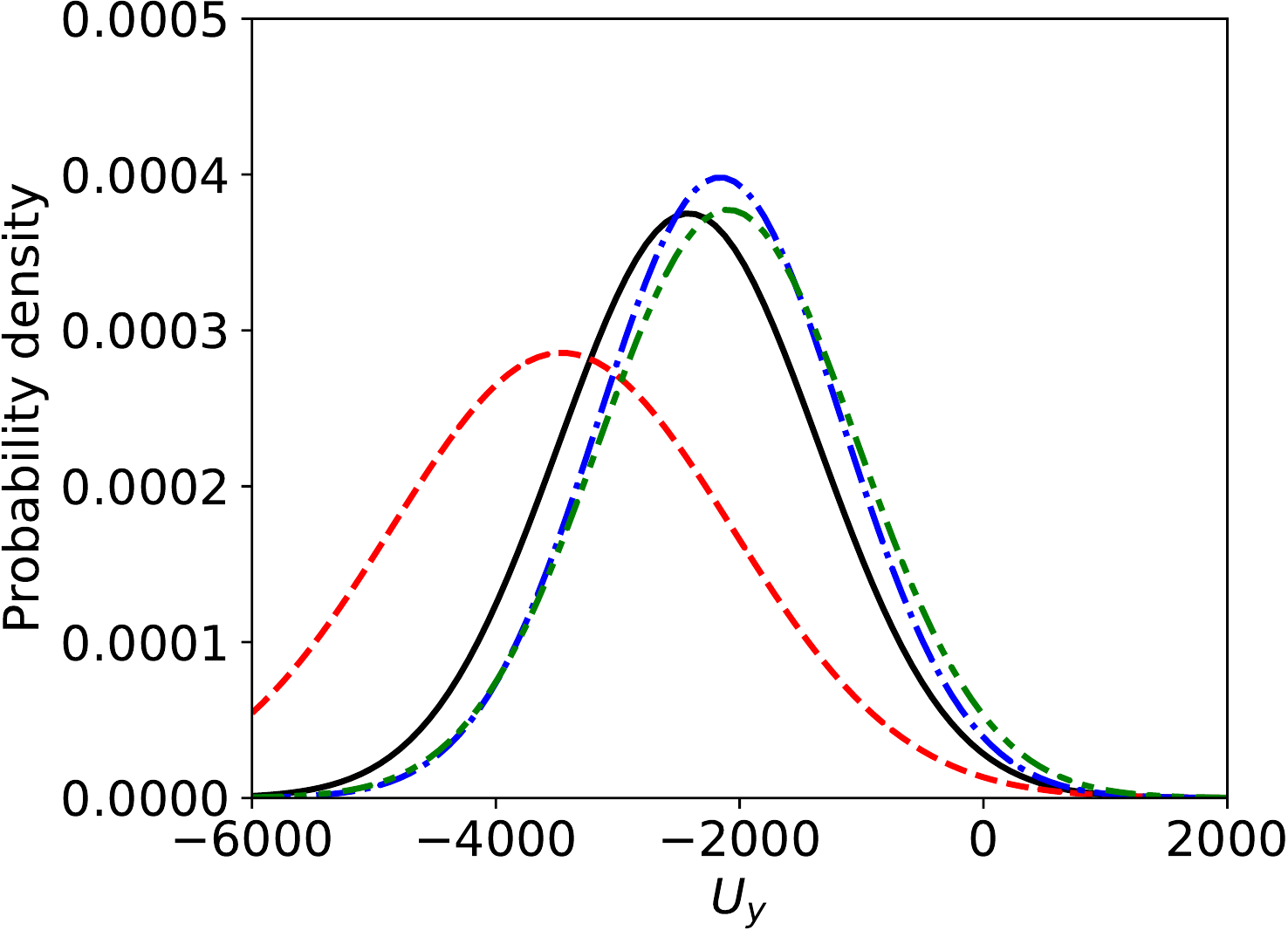}}
    \caption{The comparison of (a) horizontal velocity and (b) vertical velocity distribution of high Rayleigh number 2-D Rayleigh-B\'enard convection at the \textbf{corner} point of the training domain.}
  \label{fig:RBC-U-corner}
\end{figure}

We further studied another dataset of Rayleigh-B\'enard convection simulation at a higher Rayleigh number. More details about the Lattice Boltzmann simulation can be found in Section~\ref{sec:LBM}. As shown in Figs.~\ref{fig:RBC-U-center} and~\ref{fig:RBC-U-corner}, velocity distributions at two typical points confirm that the constrained GAN can better emulate the training data with the same training epochs. Specifically, the results at center point of the training domain is presented in Fig.~\ref{fig:RBC-U-center}, and the results at the upper right $1/4$ point (referred to as \emph{corner} point) along the diagonal of training domain are presented in Fig.~\ref{fig:RBC-U-corner}. It can be seen in Fig.~\ref{fig:RBC-U-center} that the results of standard GAN at 20 epochs are noticeably biased. Similar biased results of the standard GAN at 20 epochs can also be observed in the results at the corner point as shown in Fig.~\ref{fig:RBC-U-corner}. With 100 epochs, the results of standard GAN demonstrate much better agreement with the training data. On the other hand, the distributions of velocity at these two points from constrained GAN with only 20 epochs are much better than the results of standard GAN with the same training epochs. In addition, the results from the constrained GAN are even comparable to the results from standard GAN with 100 epochs, indicating that the mean velocity field and the TKE of the training data are reasonably captured by using constrained GAN with less computational cost. More quantitative comparisons between distributions in Figs.~\ref{fig:RBC-U-center} and~\ref{fig:RBC-U-corner} are presented in Table~\ref{tab:wasserstein} by using Wasserstein distance.

\begin{table}[htbp] 	
  \centering
  \caption{Wasserstein distance between training distribution and GAN emulated velocity distribution.
}
\label{tab:wasserstein}
\begin{tabular}{P{5.0cm} | P{2.0cm}  P{2.0cm}  P {2.0cm}  P{2.0cm}}	
  \hline
  & \multicolumn{4}{c}{Wasserstein distance} \\
  & Center $U_x$ & Center $U_y$ & Corner $U_x$ & Corner $U_y$ \\
  \hline
 Standard GAN\newline(20 epochs)  & 2455 & 2795 & 3287 & 1070 \\ 
  \hline
 Standard GAN\newline(100 epochs)  & 167 & 491 & 200 & 265 \\  
  \hline	
 Constrained GAN\newline(20 epochs)  & 463 & 1171 & 114 & 323 \\  
  \hline	
\end{tabular}
\flushleft
\end{table}

\begin{figure}[!htbp]
  \centering
  \includegraphics[width=0.3\textwidth]{field-legend}\\
  \subfloat[Training data]{\includegraphics[width=0.3\textwidth]{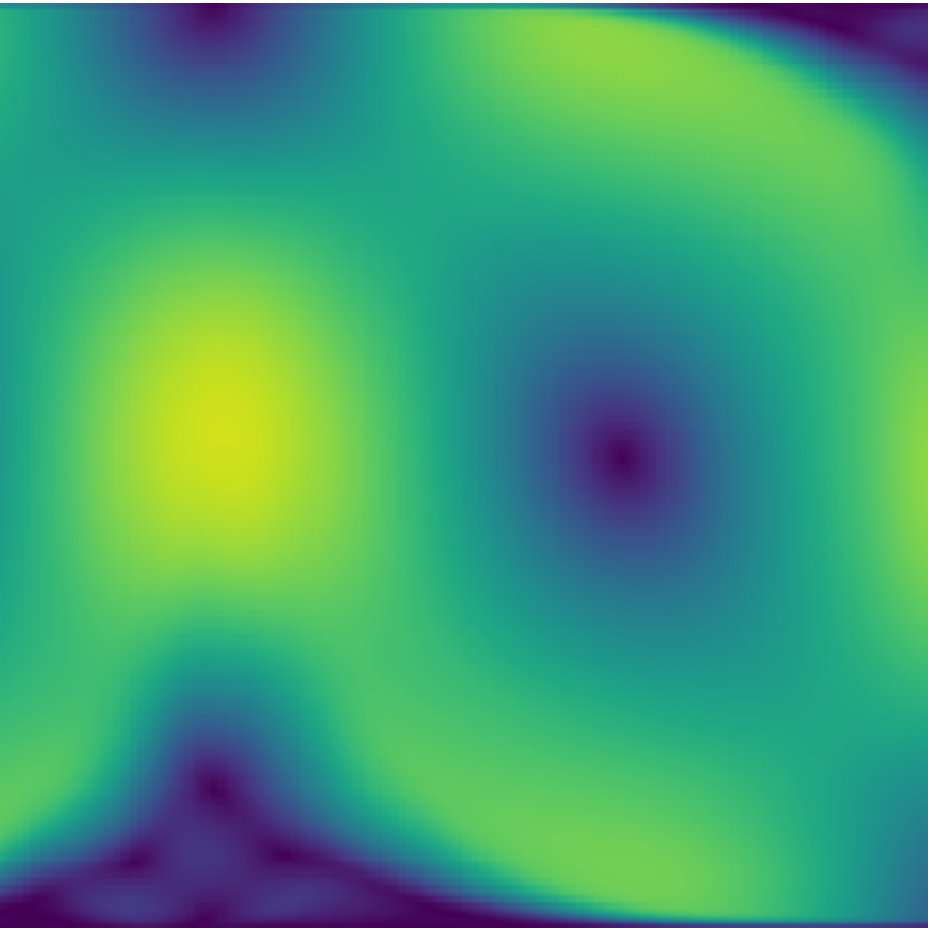}}
\hspace{0.1em}
  \subfloat[Constrained (20 epochs)]{\includegraphics[width=0.3\textwidth]{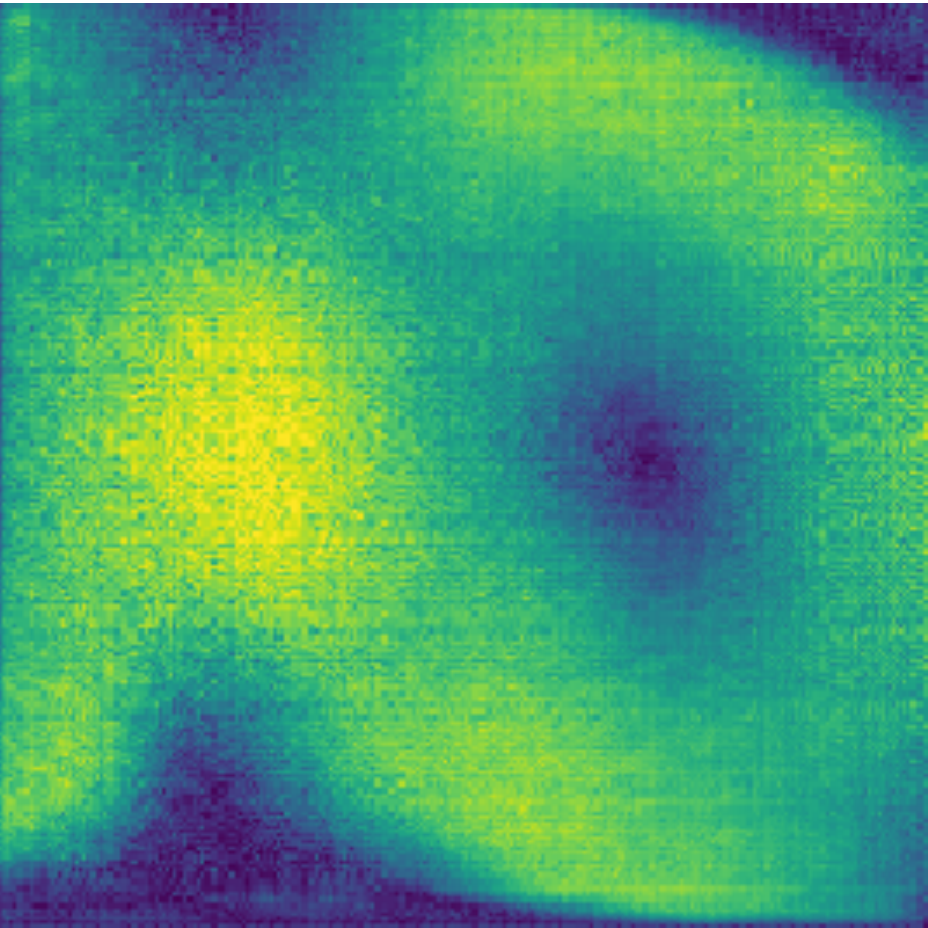}}\\
  \subfloat[Standard (20 epochs)]{\includegraphics[width=0.3\textwidth]{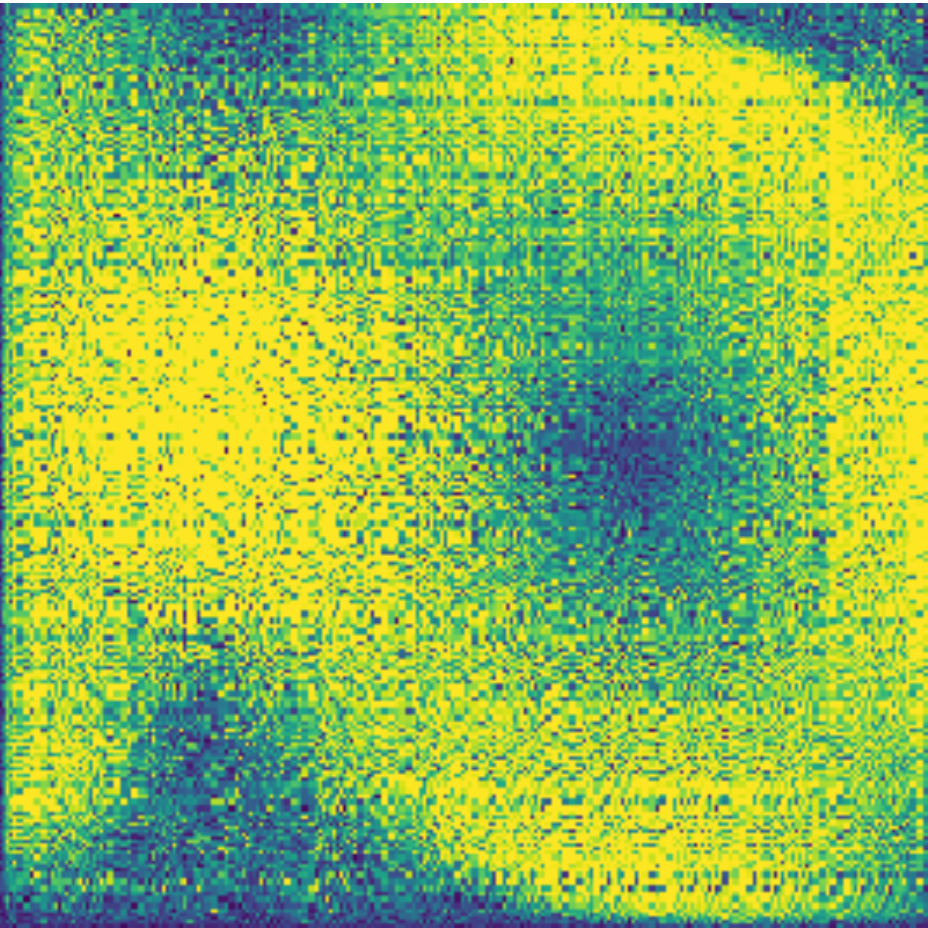}}
\hspace{0.1em}
  \subfloat[Standard (100 epochs)]{\includegraphics[width=0.3\textwidth]{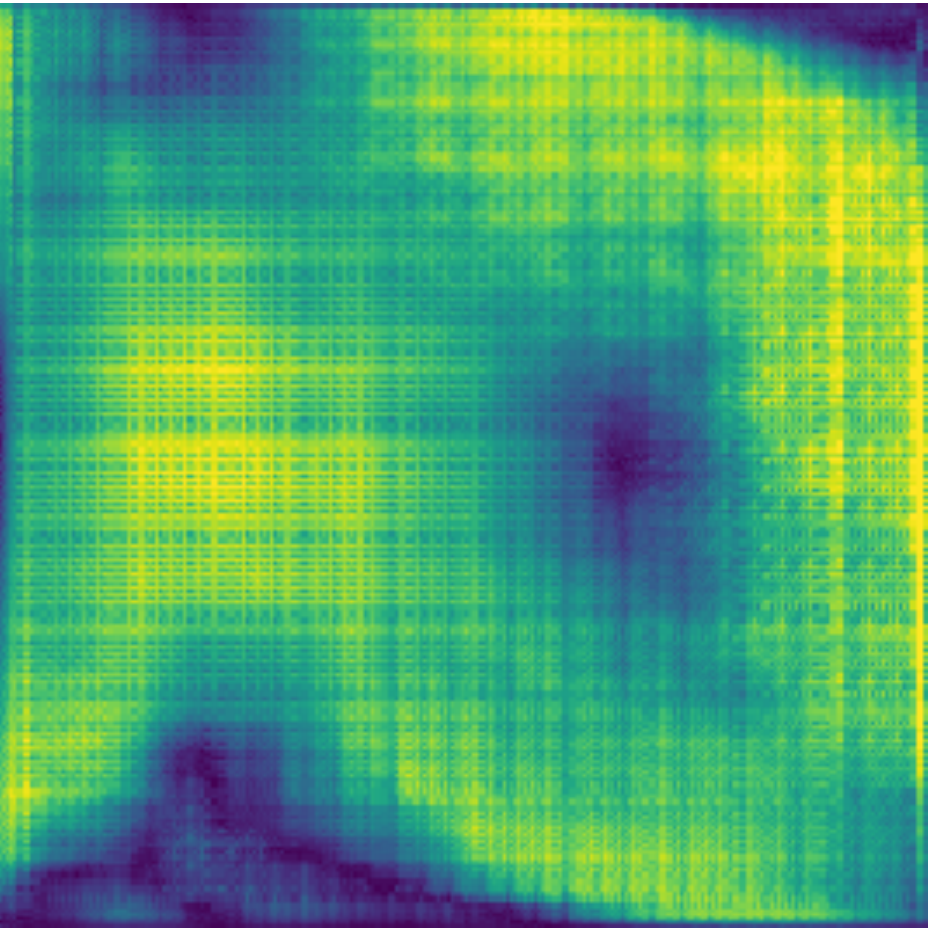}}
    \caption{The comparison of mean velocity magnitude of 2-D Rayleigh-B\'enard convection at high Rayleigh number from (a) training data, (b) constrained GAN at 20 epochs, (c) standard GAN at 20 epochs and (d) standard GAN at 100 epochs.}
  \label{fig:RBC-U-mag}
\end{figure}

The comparison of mean velocity magnitude in Fig.~\ref{fig:RBC-U-mag} shows that the standard GAN at 20 epochs noticeably overestimates the velocity magnitude across the whole domain as shown in Fig.~\ref{fig:RBC-U-mag}c, even though the circular flow pattern is qualitatively emulated. With 100 epochs, the mean velocity provided by the standard GAN is getting more similar to the training data in large scale structure. However, there are unphysical small structures existing in Fig.~\ref{fig:RBC-U-mag}d, indicating that the spatial correlation of mean velocity is not well captured by using the standard GAN. Compared to the results of the standard GAN, the result of constrained GAN in Fig.~\ref{fig:RBC-U-mag}b better emulates the training data with only 20 epochs. Although there are still some noises in small scale in Fig.~\ref{fig:RBC-U-mag}b, the noise level is much less than the result of standard GAN with 100 epochs.

\begin{figure}[!htbp]
  \centering
  \includegraphics[width=0.3\textwidth]{field-legend}\\
  \subfloat[Training data]{\includegraphics[width=0.3\textwidth]{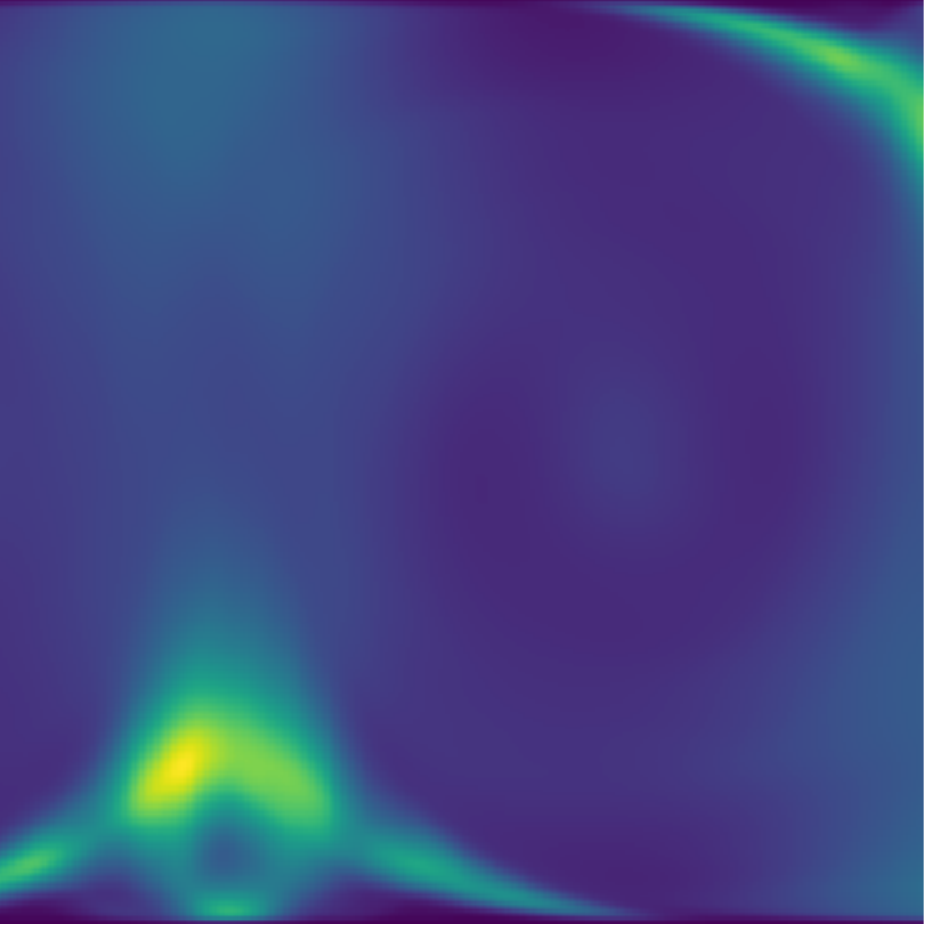}}
\hspace{0.1em}
  \subfloat[Constrained (20 epochs)]{\includegraphics[width=0.3\textwidth]{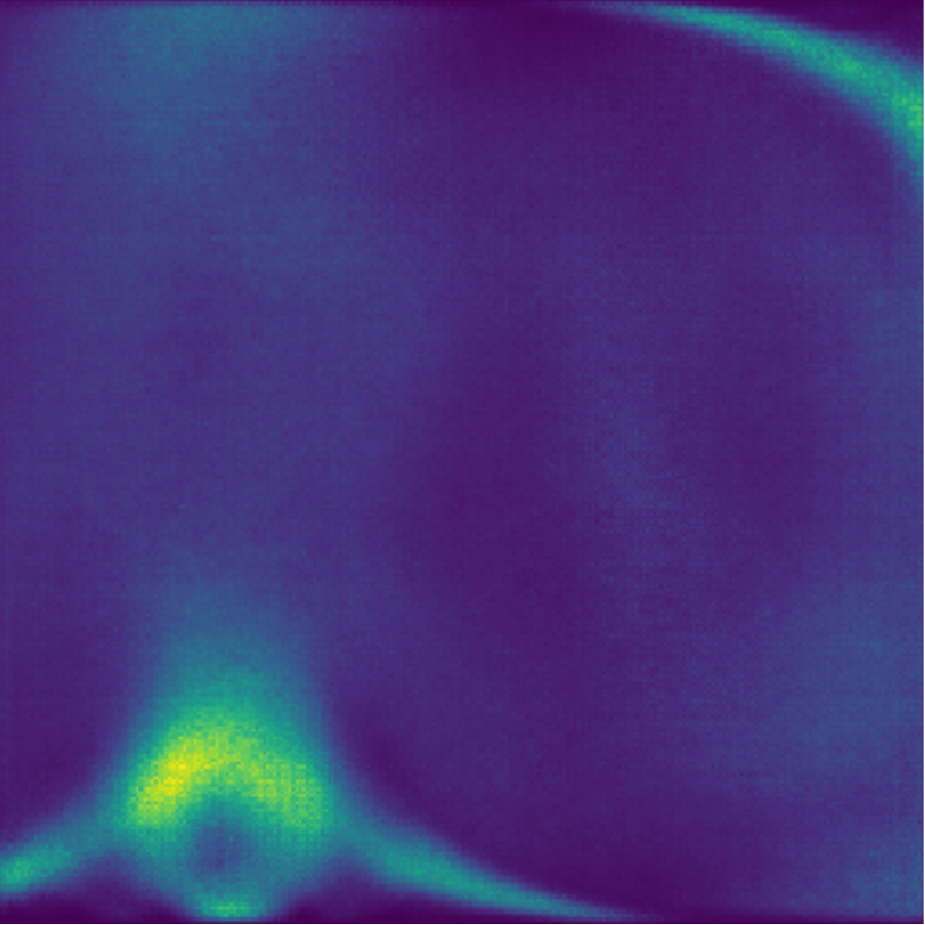}}\\
  \subfloat[Standard (20 epochs)]{\includegraphics[width=0.3\textwidth]{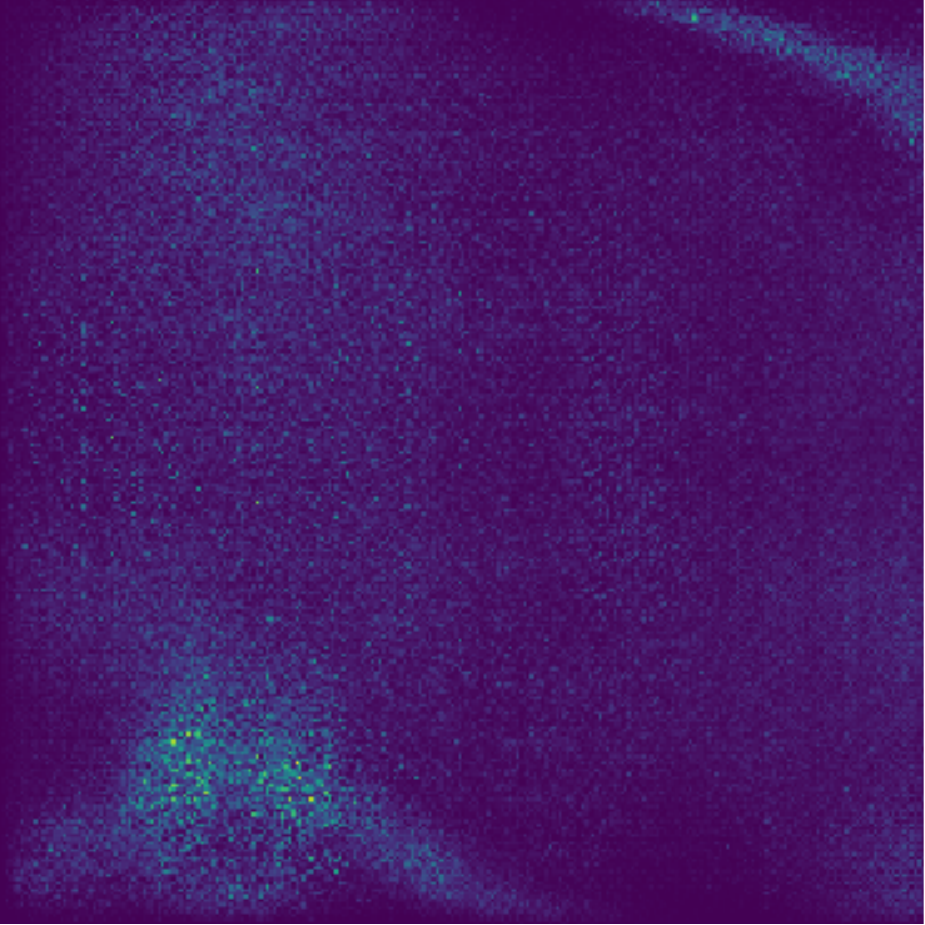}}
\hspace{0.1em}
  \subfloat[Standard (100 epochs)]{\includegraphics[width=0.3\textwidth]{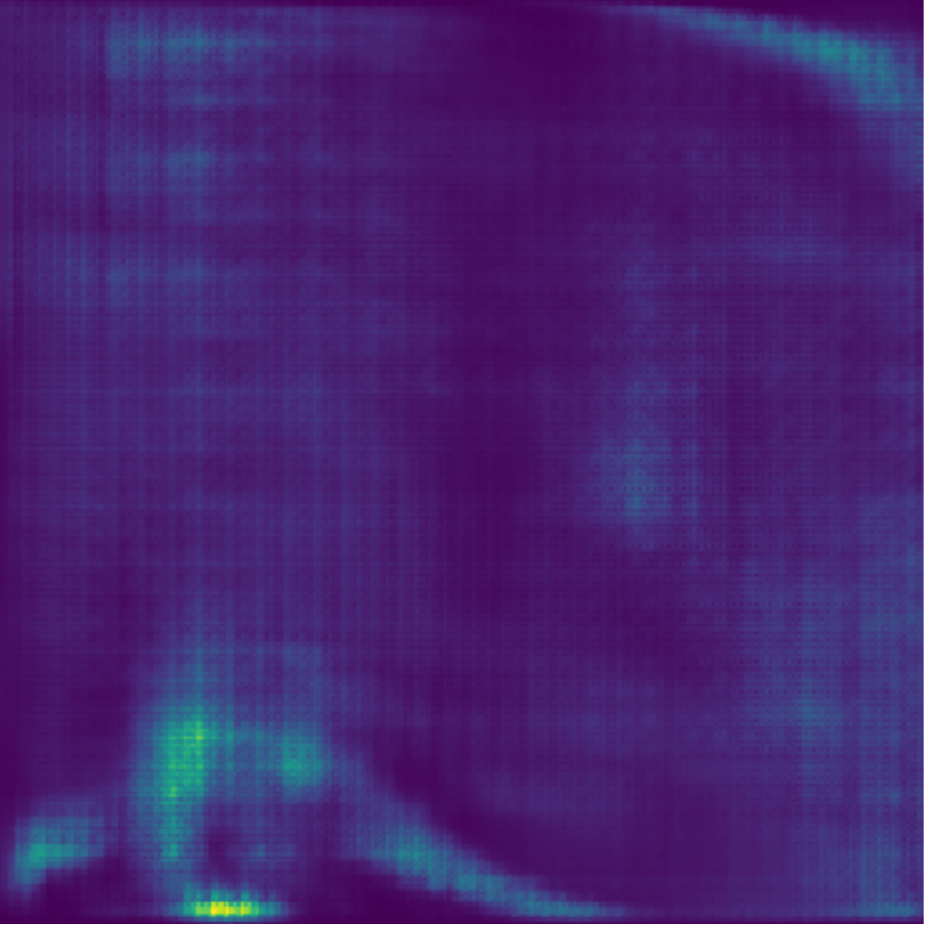}}
    \caption{The comparison of turbulent kinetic energy of 2-D Rayleigh-B\'enard convection at high Rayleigh number from (a) training data, (b) constrained GAN at 20 epochs, (c) standard GAN at 20 epochs and (d) standard GAN at 100 epochs.}
  \label{fig:RBC-tke}
\end{figure}

The comparison of the TKE fields as shown in Fig.~\ref{fig:RBC-tke} also demonstrates the better performance of the constrained GAN. It should be noted that the result in Fig.~\ref{fig:RBC-tke}d from the standard GAN involves 100 epochs, while the result in Fig.~\ref{fig:RBC-tke}b from the statistical constrained GAN only involves 20 epochs. With less training epochs, the result of the statistical constrained GAN still shows better agreement with the training data. Specifically, the spatial pattern of the training data around the left bottom corner region is much better captured by the constrained GAN. Also, improvements can be observed at other regions across the whole domain as shown in Fig.~\ref{fig:RBC-tke}. The results in Figs.~\ref{fig:RBC-U-mag} and~\ref{fig:RBC-tke} together demonstrate that the constrained GAN is a more practical tool for preserving important statistics (e.g., mean and second-order moment) when emulating complex systems.

\begin{figure}[!htbp]
  \centering
  \hspace{2em}\includegraphics[width=0.6\textwidth]{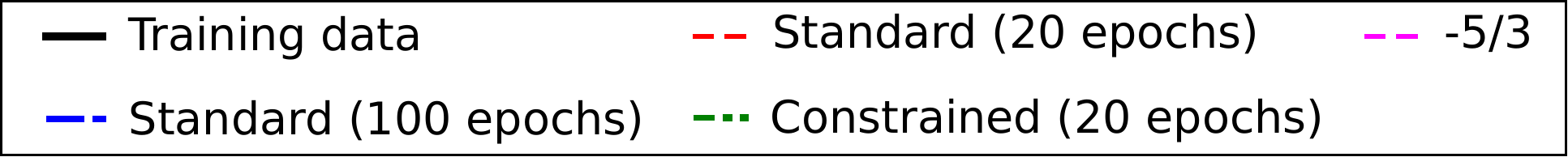}
  \includegraphics[width=0.6\textwidth]{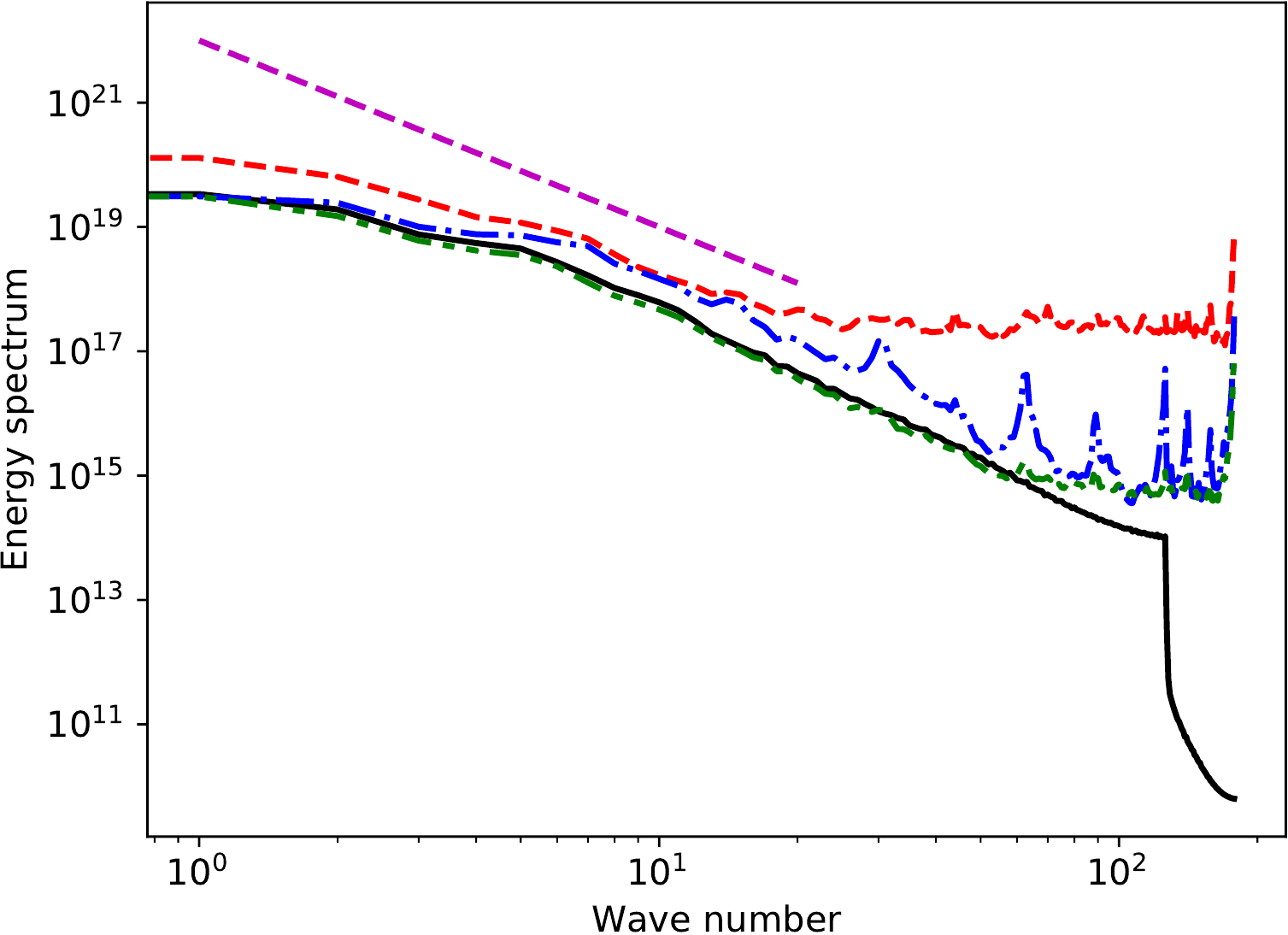}
    \caption{Comparison of turbulent kinetic energy spectrum of 2-D Rayleigh-B\'enard convection with a high Rayleigh number.}
  \label{fig:RBC-fine-spectrum}
\end{figure}

We also studied the spectrum of TKE in Fig.~\ref{fig:RBC-fine-spectrum}. It can be seen that the result of standard GAN at 20 epochs overestimates energy spectrum across the whole range of wave numbers. With 100 training epochs, the result of the standard GAN better captures the energy spectrum for relatively small wave numbers (wave number is less than 5), but still shows noticeable difference from training data for large wave numbers (small scale structure in space). In addition, the noise level increases with more training epochs by using the standard GAN, indicating that asymptotically improvement of performance may not be achieved with more training epochs. On the other hand, the result of the constrained GAN at 20 epochs successfully captured the energy spectrum of the training data for most wave numbers, with only an exception of very large wave numbers ($\sim 10^2$). 

\section{Conclusion}
Recently, generative adversarial networks (GANs) have shown great potential of emulating and even predicting the solutions of PDEs-governed physical systems by training on some high-fidelity simulation data. However, it is known that GANs are much more difficult to train than regular neural networks, preventing its application to more complex and realistic physical systems. In this work, we introduce a statistical constrained approach to improve the robustness of training GANs. Specifically, the difference between the covariance structures of the generated samples and the training data is quantified and incorporated into the original loss function of the generator in GANs as a penalty term. The performance of the constrained GAN is evaluated against the standard GANs by studying the Gaussian random field and the Rayleigh--B\'enard convection. The results demonstrated that the proposed statistical regularization leads to more robust training than standard GANs. Even though similar quality of results can be achieved by carefully tuning the hyper-parameters of standard GANs, the constrained GAN significantly reduced (by up to 80\%) training cost to reach the solution with similar or even better quality. With the growth of high-fidelity simulation databases of physical systems, this work has a great potential of being an alternative to the explicit modeling of closures or parameterizations for unresolved physics. By better preserving high-order statistics from existing high-fidelity simulations, this work potentially leads to a class of promising application of GANs in simulating chaotic physical systems, e.g., turbulence or Earth's climate.

\section*{Acknowledgments}
This research used resources of the National Energy Research Scientific Computing Center, a DOE Office of Science User Facility supported by the Office of Science of the U.S. Department of Energy under Contract No. DE-AC02-05CH11231.

\end{document}